\def\be{\begin{equation}}
\def\ee{\end{equation}}
\def\bea{\begin{eqnarray*}}
\def\eea{\end{eqnarray*}}

\def\mnras{MNRAS}

\def\apjl{ApJ}
\def\apj{ApJ}
\def\apjs{ApJS}
\def\pasp{PASP}
\def\nat{Nature}
\def\physrep{Phys.~Rep.}   
\def\aj{AJ}

\documentclass[useAMS,usenatbib]{mn2e}
\usepackage{epsfig}
\usepackage{subfigure}
\usepackage{graphicx}
\usepackage{flafter}
\usepackage{lscape}
\usepackage{amsmath}
\newcommand{\Eq}{{Equation~}}
\newcommand{\beq}{\begin{equation}}
\newcommand{\eeq}{\end{equation}}
\newcommand\T{\rule{0pt}{2.6ex}}
\newcommand\B{\rule[-1.2ex]{0pt}{0pt}}

\def\lsim{\mathrel{\lower2.5pt\vbox{\lineskip=0pt\baselineskip=0pt
           \hbox{$<$}\hbox{$\sim$}}}}
\def\gsim{\mathrel{\lower2.5pt\vbox{\lineskip=0pt\baselineskip=0pt
           \hbox{$>$}\hbox{$\sim$}}}}

\input epsf

\title[AzTEC/SHADES SMG Survey] 
{AzTEC Half Square Degree Survey of the SHADES Fields -- I. Maps, Catalogues, and Source Counts}
\author[J.E. Austermann et al.]{%
J.E.~Austermann$^{1,2}$, J.S.~Dunlop$^{3,4}$, 
T.A.~Perera$^1$, K.S.~Scott$^1$, G.W.~Wilson$^1$,
\newauthor
I.~Aretxaga$^5$,
D.H.~Hughes$^5$,
O.~Almaini$^6$,
E.L.~Chapin$^4$, 
S.C.~Chapman$^7$,  
\newauthor
M.~Cirasuolo$^3$,
D.L.~Clements$^8$,
K.E.K.~Coppin$^9$, 
L.~Dunne$^6$, 
S.~Dye$^{10}$, 
S.A.~Eales$^{10}$,  
\newauthor
E.~Egami$^{11}$,
D.~Farrah$^{12,13}$, 
D.~Ferrusca$^{5}$,
S.~Flynn$^{10}$,
D.~Haig$^{10}$,
M.~Halpern$^4$,  
\newauthor
E.~Ibar$^{14}$,
R.J.~Ivison$^{3,14}$, 
E. van Kampen$^{15}$,
Y.~Kang$^{16}$,
S.~Kim$^{16}$, 
C.~Lacey$^{9}$,
\newauthor
J.D.~Lowenthal$^{17}$,
P.D.~Mauskopf$^{10}$,
R.J.~McLure$^3$,
A.M.J.~Mortier$^3$, 
M.~Negrello$^{18}$, 
\newauthor  
S.~Oliver$^{12}$,
J.A.~Peacock$^3$, 
A.~Pope$^{19}$, 
S.~Rawlings$^{20}$,
G.~Rieke$^{11}$, 
I.~Roseboom$^{12}$
\newauthor
M.~Rowan-Robinson$^8$,
D.~Scott$^4$, 
S.~Serjeant$^{16}$, I.~Smail$^9$,
A.M.~Swinbank$^9$, 
\newauthor
J.A.~Stevens$^{21}$, 
M.~Velazquez$^{5}$,
J.~Wagg$^{22}$,
M.S.~Yun$^1$.
\\
$^1$Department of Astronomy, University of Massachusetts, Amherst, MA 01003, USA\\
$^2$Center for Astrophysics and Space Astronomy, University of Colorado, Boulder, CO 80309, USA\\
$^3$Scottish Universities Physics Alliance (SUPA), 
Institute for Astronomy, University of Edinburgh, 
Royal Observatory, Edinburgh
EH9 3HJ\\
$^4$ Department of  Physics and  Astronomy, University of British Columbia, 
6224 Agricultural Rd., Vancouver, B.C., V6T 1Z1, Canada\\
$^5$ Instituto Nacional de Astrof\'{\i}sica, \'Optica y Electr\'onica 
(INAOE), Aptdo. Postal 51 y 216, 72000 Puebla, Mexico\\
$^6$ School of Physics and Astronomy, University of Nottingham,
University Park, Nottingham NG7 2RD\\
$^7$ Institute of Astronomy, University of Cambridge, Madingley Road,
 Cambridge, CB3 0HA\\
$^8$ Imperial College, London, Blackett Lab, Prince Consort Road, London
 SW7 2BW\\
$^9$ Institute for Computational Cosmology, Durham University, 
South Road, Durham DH1 3LE\\
$^{10}$ School of Physics \& Astronomy, Cardiff University, Queens Buildings, The Parade, 
Cardiff CF24 3AA\\
$^{11}$ Department of Astronomy, University of Arizona, 
933 North Cherry Avenue, Tucson, AZ, 85721-0065, USA\\
$^{12}$ Astronomy Centre, University of Sussex, Falmer, Brighton BN1 9QH\\
$^{13}$ Department of Astronomy, Cornell University, Space Sciences Buildin
g, Ithaca, NY 14853, USA\\
$^{14}$ UK Astronomy Technology Centre, Royal Observatory, Edinburgh EH9 3HJ\\
$^{15}$ ESO, Karl-Schwarzschild-Str. 2, D-85748 Garching bei M\"unchen, Germany\\
$^{16}$ Astronomy \& Space Science Department, Sejong University, Seoul, South Korea.\\
$^{17}$ Department of Astronomy, Smith College, Northampton, MA 01063, USA.\\
$^{18}$ Department of Physics and Astronomy, Open University, 
Milton Keynes, MK7 6AA\\
$^{19}$ National Optical Astronomy Observatory, 950 North Cherry Avenue, 
Tucson, AZ, 85719, USA\\
$^{20}$ Department of Astrophysics, Denys Wilkinson Building, Keble Road
, Oxford OX1 3RH\\
$^{21}$ Centre for Astrophysics Research, Science and Technology Research
Centre, University of Hertfordshire, College Lane, Herts AL10 9AB\\
$^{22}$ National Radio Astronomy Observatory, PO Box O, Socorro, NM, USA 87801\\
\\
\\
\\
\\
\\
\\
\\
\\
\\
\\
\\
\\
\\
\\
\\
}

\begin{document}

\date{}

\pagerange{} \pubyear{}

\maketitle

\label{firstpage}

\begin{abstract}
We present the first results from the largest deep extragalactic
millimetre-wavelength survey undertaken to date. These results are
derived from maps covering over 0.7\,deg$^2$, made at $\lambda =
1.1$\,mm, using the AzTEC continuum camera mounted on the James Clerk
Maxwell Telescope. The maps were made in the two fields originally
targeted at $\lambda = 850$\,$\micron$ with the Submillimetre
Common-User Bolometer Array (SCUBA) in the SHADES project, namely the
Lockman Hole East (mapped to a depth of 0.9--1.3~mJy rms) and the
Subaru XMM Deep Field (mapped to a depth of 1.0--1.7~mJy rms). The
wealth of existing and forthcoming deep multi-frequency data in these
two fields will allow the bright mm source population revealed by
these new wide-area 1.1\,mm images to be explored in detail in
subsequent papers. Here we present the maps themselves, a catalogue of
114 high-significance sub-millimetre galaxy detections, and 
a thorough statistical analysis leading to the most robust determination
to date of the 1.1\,mm source number counts. 
These new maps, covering an area $\sim3$ times greater than the SCUBA SHADES maps, 
currently provide the largest sample of cosmological volumes of the high-redshift Universe 
in the mm or sub-mm.  
Through careful comparison, we
find that both the COSMOS and GOODS North fields, also imaged with
AzTEC, contain an excess of mm sources over the new 1.1\,mm
source-count baseline established here. In particular, our new
AzTEC/SHADES results indicate that very luminous high-redshift dust
enshrouded starbursts ($S_{\rm{1.1mm}}>3$\,mJy) are 25--50~per cent less common
than would have been inferred from these smaller surveys, thus
highlighting the potential roles of cosmic variance and clustering in
such measurements.  We compare number count predictions from recent models of
the evolving mm/sub-mm source population to these SMG surveys, which
provide important constraints for the ongoing refinement of
semi-analytic and hydrodynamical models of galaxy formation, and find
that all available models over-predict the number of bright
sub-millimetre galaxies found in this survey.
\end{abstract}



\section{Introduction}
\label{sec:intro}

In the last two decades, surveys in the far-infrared and sub-millimetre 
have revolutionised our understanding of galaxy evolution in the 
high-redshift Universe. These surveys, primarily at wavelengths around
850--$\micron$ 
(e.g. \citealp{smail97,hughes98,barger98,scott02,borys03,pope06,coppin06,greve09}), 
1100--$\micron$ \citep{laurent05,scott08,perera08}, 
and 1200--$\micron$ \citep[e.g.][]{greve04,bertoldi07,greve08} 
have shown that the 
contribution to the comoving IR energy density from sub-mm bright 
galaxies (SMGs) increases by approximately three orders of magnitude 
in going from the local Universe to $z\gsim 1$, and that $z\gsim 1$ 
SMGs are responsible for a significant portion 
the extragalactic infrared background light \citep[e.g.][]{scott08,serjeant08}.
Initial followup studies have identified optical counterparts 
\citep[e.g.][]{dye08,clements08} and measured redshifts 
\citep[e.g.][]{chapman05,aretxaga07} of many SMGs. 
Further studies have shown that SMGs 
harbour very high rates of star formation, often accompanied by 
significant AGN activity 
\citep[e.g.][]{alex05,kovacs06,coppin08,menendez-delmestre07,pope08} and 
that at least some are involved in ongoing mergers \citep{farrah02,chap03}, 
implying that SMGs signpost massive galaxy assembly in the high redshift 
Universe. Reviews of their properties can be found in \citet{blain02} and 
\citet{lons06}.

On a fundamental level, SMGs represent the efficient transformation of 
free baryons into stars and black holes. Recent work has therefore focused 
on understanding how SMGs relate to the cosmological evolution of the total 
and baryonic mass density. Evidence suggests that this relationship is complex, 
with an intricate dependence on variables such as redshift, local 
environmental richness, and halo merger history. Accordingly, observations 
must find SMGs across a wide range of environments and redshifts. 
This has traditionally proven difficult; SMGs are easy to find across wide 
redshift ranges due to the favourable k-correction at sub-mm wavelengths, 
but hard to find across wide ranges in environment due to the inability of 
most sub-mm bolometer arrays to efficiently map large areas of sky. 
Constraints on SMG number counts have thus been limited by both sample size 
and cosmic variance, we have found relatively few of the brightest and 
rarest SMGs, and constraints on SMG clustering -- an important tool in 
relating SMGs to the underlying dark matter distribution -- are weak 
(\citealp[e.g.][]{blain04,vankampen05,chapman08}, see also 
\citealp{farrah06,mag07}). 
As a result, many studies have adopted a two-pronged 
approach; using modest sized blank field surveys to constrain the properties 
of the general SMG population, combined with targeted surveys of clusters to 
probe the properties of SMGs in the highest density regions 
\citep[e.g.][]{stevens03,greve07,priddey08,austermann08,tamura09}. 
Even this approach has drawbacks though, as it requires a pre-existing 
cluster catalogue extending to redshifts significantly in excess of unity, 
where clusters are difficult to find.

This situation has recently been improved by both the combined analysis of 
multiple surveys \citep[e.g.][]{scott06} and by the advent of larger area 
sub-mm surveys such as the 850--$\micron$ SCUBA/SHADES survey 
\citep{coppin06}, which mapped 0.2\,deg$^2$ to depths of 
$\sigma_{850} \sim 2$\,mJy. In this paper, we present a further step forward 
in understanding the SMG population through the 1100--$\micron$ AzTEC/SHADES 
survey, which covers 0.5\,deg$^2$ to depths of $\sigma_{1100} \sim 1$\,mJy and
over 0.7\,deg$^2$ in total. 
This survey dramatically improves our understanding of the 1100--$\micron$ 
blank-field population, with previous 1100--$\micron$ surveys being smaller 
in area \citep[e.g.][]{perera08}, shallower \citep[e.g.][]{laurent05}, 
or containing known biased regions in their survey volumes 
\citep[e.g.][]{austermann08,tamura09}.

This paper is organised as follows. 
Details of the AzTEC observations are given in Section~\ref{sec:observations}, 
and we present the 1100--$\micron$ maps and source catalogues in Section~\ref{sec:maps}. 
Detailed constraints on the 1100--$\micron$ blank-field number counts 
are derived in Section~\ref{sec:counts}, 
and we compare these results to those derived from other surveys 
and models in Section~\ref{sec:discussion}. 
Finally, our conclusions are summarised in Section~\ref{sec:conclusions}.
We assume a flat $\Lambda$CDM cosmology with
$\Omega_M = 0.3$, $\Omega_{\Lambda} = 0.7$, and
$H_0 = 73$~km~s$^{-1}$~Mpc$^{-1}$. 
This paper is the first in a series of papers using the AzTEC/SHADES maps 
and catalogues to study the sub-millimetre population of galaxies.


\section{Observations}
\label{sec:observations}

We have completed the SCUBA Half-Degree Extragalactic Survey
\citep[SHADES; ][]{mortier05} by mapping over one-half square degree of sky 
using the AzTEC 1.1--mm camera \citep{wilson08a} mounted on the
15-metre James Clerk Maxwell Telescope (JCMT).  The AzTEC/JCMT system
results in a Gaussian beam with $\textrm{FWHM}\approx18$\,arcsec.  The
SHADES survey is split between the Lockman Hole East field (LH;
10$^\textrm{h}$52$^\textrm{m}$, $+$57$^{\circ}$00$^\prime$) and the
Subaru/XMM-Newton Deep Field (SXDF; 02$^\textrm{h}$18$^\textrm{m}$,
$-$05$^{\circ}$00$^\prime$).  AzTEC has mapped over 0.7\,deg$^2$ to
1.1\,mm depths of 0.9--1.7\,mJy between the LH and SXDF fields,
including the central 0.13 and 0.11\,deg$^2$, respectively, mapped by
SCUBA at 850\,$\micron$
\citep{coppin06}.  All AzTEC/SHADES observations were carried out
between November 2005 and January 2006, with over 180 hours of
telescope time dedicated to this project, including all overheads.

\begin{figure}
\epsfig{file=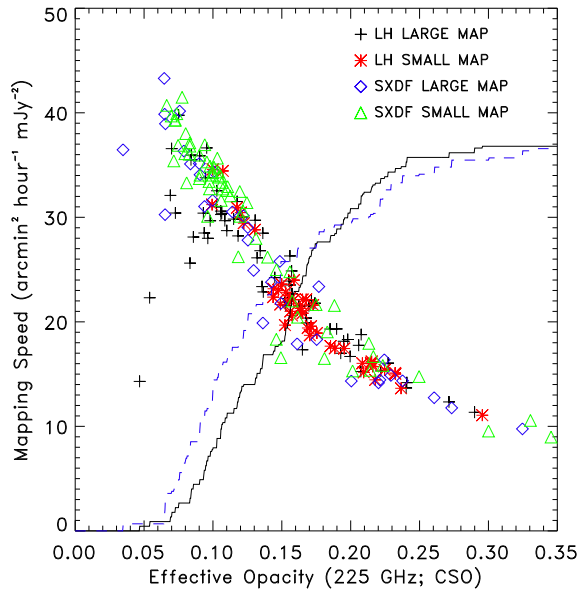, width=1.0\hsize,angle=0}
\caption{Achieved mapping speeds for the various AzTEC/SHADES 
observations.  Mapping speeds do not include overheads and are 
calculated as defined in \citet{wilson08a}.
The cumulative distributions of observation time going into 
the final AzTEC/LH (92 hours) and AzTEC/SXDF (61 hours) maps are plotted as the 
solid and dashed curves, respectively, and normalised to an arbitrary 
value on the y-axis.   
}
\label{fig:mapping_speed}
\end{figure}

The AzTEC/SHADES observing strategy is similar to that used for other
AzTEC blank-field surveys at the JCMT and is described in detail in
previous publications \citep{scott08,perera08}.  All observations were
made while scanning the telescope in elevation in a raster pattern
(see \citealp{wilson08a}).  Initially, the AzTEC/SHADES observations
were made as small 15\,arcmin $\times$ 15\,arcmin mosaic maps with
scan speeds of 90--120\,arcsec\,s$^{-1}$.  Later observations were
extended to cover an entire field in one continuous observation of
size 35\,arcmin $\times$ 35\,arcmin, which served to reduce
observational overheads.  Faster scan speeds of
180--220\,arcsec\,s$^{-1}$ were used for these longer scans, which
increased the effective sensitivity of the observations due to the
corresponding reduction in residual atmospheric noise at the higher
temporal frequencies \citep{wilson08a}.  After full reduction, the
larger maps with faster scan-speeds have an observing efficiency of
$\sim$ 150~per cent, relative to the smaller maps.  In the end, 46
(63) mosaic maps and 65 (34) full maps were used to create the final
LH (SXDF) map.  These observations were performed over a wide range of
atmospheric conditions and elevations.  
Figure~\ref{fig:mapping_speed} shows both
the achieved mapping speeds and the cumulative distribution of observation 
time as a function of effective opacity at 225\,GHz.

Nightly overhead observations included focusing, load curves, beam
maps and pointing observations, all of which are described in the
AzTEC instrument paper \citep{wilson08a}.  Pointing observations of
bright point sources (typically $> 1$\,Jy) that lie near the science
field being targeted were made every two hours.  These measurements
provide small corrections to the JCMT pointing model and are applied
using a linear interpolation between the nearest pointing measurements
taken before and after each science observation.  Flux calibration is
performed as described in \citet{wilson08a} using the nightly load
curves and beam maps of our primary calibration source, Uranus.  The
error in flux calibration is estimated to be 6--13~per cent on an
\textit{individual} observation \citep{wilson08a}.  The actual error
in the final co-added AzTEC/SHADES maps, which comprise observations
spanning many nights and calibrations, will be smaller, assuming the
calibration uncertainty is randomly distributed.  These individual
error estimates do not include the systematic 5~per cent absolute
uncertainty in the flux density of Uranus \citep{griffin93}.

\begin{figure*}
\centering
\subfigure[Lockman Hole East (LH)]
{
\label{fig:map_a}
\includegraphics[width=8.5cm]{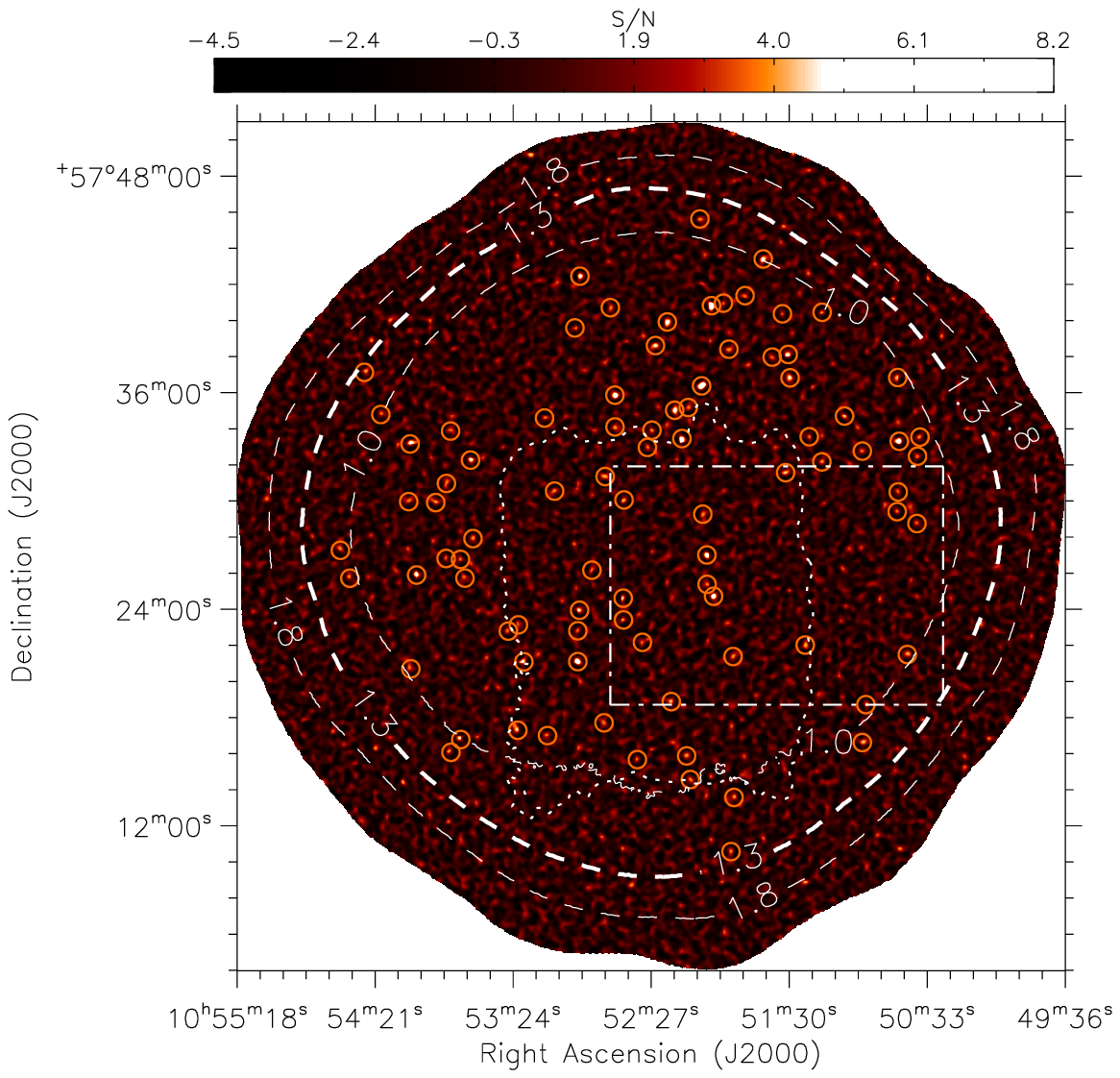}
}
\subfigure[Subaru/XMM-Newton Deep Field (SXDF)]
{
\label{fig:map_b}
\includegraphics[width=8.5cm]{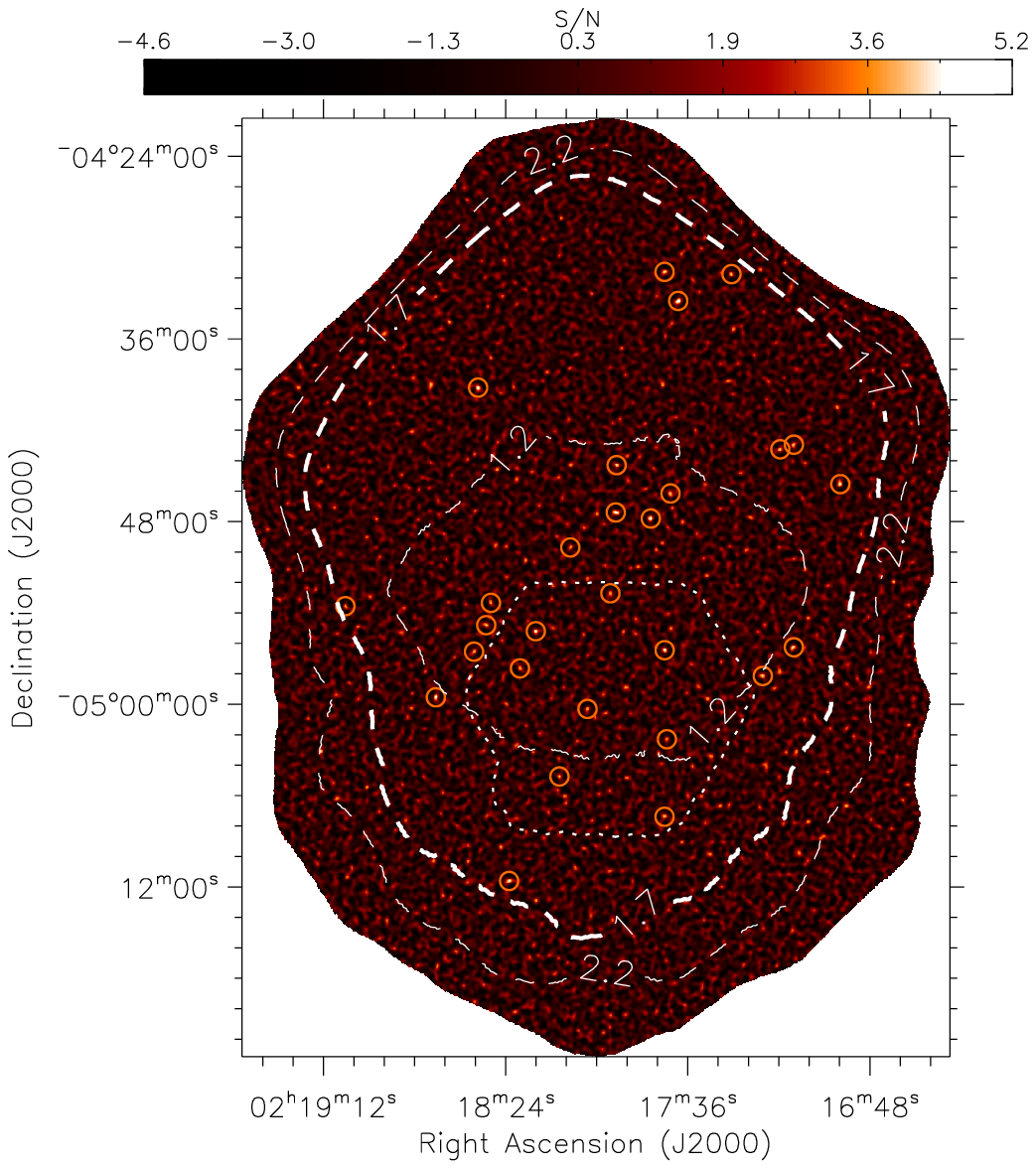}
}
\caption{AzTEC Signal-to-Noise maps of the Lockman Hole East 
field and SXDF, each observed with an AzTEC/JCMT beamsize of
$\textrm{FWHM}\approx 18$\,arcsec.  Maps are displayed at different
scales.  The most significant source candidates, as defined in the
text and listed in
Tables~\ref{table:lh_sources}--\ref{table:sxdf_sources}, are circled.
The observing pattern results in maps that are deepest in the centre,
with noise increasing towards the edges.  Reference contours show the
distribution of 1$\sigma$ depth, in mJy.  The maximal extent
($\sigma_{850} \le 6$\,mJy) of the overlapping 850--$\micron$
SCUBA/SHADES survey is depicted by the dotted curves.  (a) AzTEC/LH
$S/N$ map with 0.45\,deg$^2$ total displayed area.  For analysis
purposes, the survey area is trimmed to the region of relatively
uniform coverage (thick 1.3\,mJy contour), resulting in a
0.31\,deg$^2$ map with noise levels 0.9--1.3\,mJy.  For comparison in
Section~\ref{sec:discussion}, the approximate size of the similar
depth 1.1--mm AzTEC/GOODS-N survey \citep[0.068\,deg$^2$;
][]{perera08} is represented by the dash-dotted rectangle; however,
these surveys do not actually overlap on the sky.  (b) AzTEC/SXDF
$S/N$ map covering a total of 0.59\,deg$^2$, with the trimmed analysis
region (thick 1.7\,mJy contour) covering 0.37\,deg$^2$ with noise
levels 1.0--1.7\,mJy.  }
\label{fig:maps}
\end{figure*}


\section{Maps and Catalogues}
\label{sec:maps}

In this section we describe the methods used to construct the 1.1--mm
maps and source catalogues.  We test the astrometry and calibration of
our maps against complementary radio data.  We also describe expanded
and improved methods for estimating and correcting for flux biases
inherent to these surveys and test these estimates against
simulations.

\subsection{\label{ssec:mapmaking}Mapmaking}

The time streams of each observation are cleared of intermittent
spikes (e.g. cosmic-ray events, instrumental glitches) and have the
dominant atmospheric signals removed using the techniques described in
\citet{scott08}.  Each observation is then mapped to a 3\,arcsec $\times$
3\,arcsec grid in RA-Dec that is tangent to the celestial sphere at
(10$^\textrm{h}$51$^\textrm{m}$59$^\textrm{s}$,
$+57^{\circ}$21$^\prime$43$^{\prime\prime}$) for Lockman Hole and
(02$^\textrm{h}$18$^\textrm{m}$01$^\textrm{s}$,
$-04^{\circ}$59$^\prime$54$^{\prime\prime}$) for SXDF.  These are the
same pixel sizes and tangent points used for the SCUBA/SHADES
850--$\micron$ maps, allowing for straightforward comparison of maps
in upcoming SHADES publications.  All observations are then `co-added'
on the same grid to provide a weighted-average signal map and weight
map for each field.

In parallel, we pass a simulated point source (as defined through beam
map observations) through the same algorithms to trace and record the
effective PSF, or `point source kernel', in our final maps.  We also
create five noise-only map realisations of each observation by
jack-knifing (randomly multiplying by $1$ or $-1$) each scan (5--15
seconds of data) of the time stream.  This process works to remove any
astronomical signal while preserving the dominant noise properties in
the map\footnote{Jack-knifing also removes confused astronomical signal, which
can sometimes be considered a source of noise. However, confusion noise is not significant 
for maps of this depth and beamsize (see Section~\ref{ssec:nc_simulations}.), as confirmed
through simulation.}. 
The resulting jack-knifed maps are dominated by residual
atmospheric contamination and detector noise.  One-hundred fully
co-added noise maps are then created by randomly selecting a noise
realisation for each observation and calculating the weighted average
in the same manner used to create the co-added signal maps.  Because 
the atmospheric contamination and the detector noise are uncorrelated
amongst the full set of maps, the resulting co-added noise maps
are, like the underlying noise in the signal map, 
extremely Gaussian.

\begin{figure}
\epsfig{file=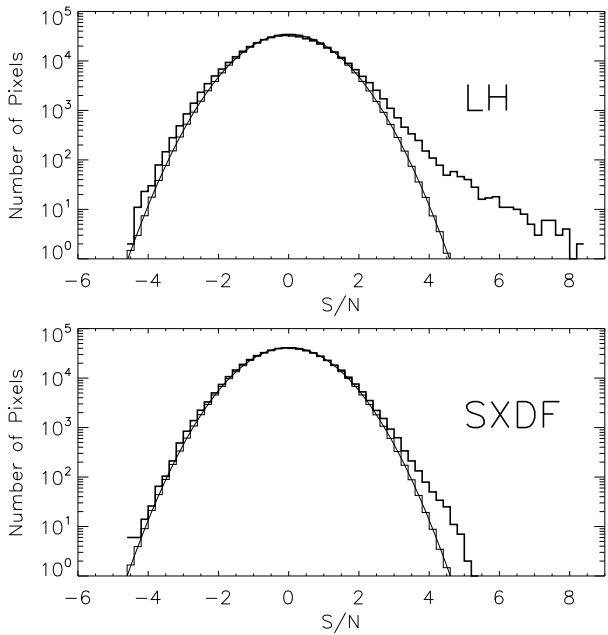, width=1.0\hsize,angle=0}
\caption{Pixel $S/N$ histograms of the LH and SXDF maps (thick histograms)
and average of their respective noise map realisations (thin histograms).  
The noise-only maps are well described as a Gaussian (smooth curve), 
while the signal maps are distorted by the presence of sources.
Sources affect both the positive and negative flux distribution 
due to the zero-mean (AC-coupled) nature of these AzTEC maps.
The distortions caused by sources are more apparent in LH due to the map's
lower average noise level compared to SXDF (Figure~\ref{fig:maps}). 
}
\label{fig:hist}
\end{figure}

An optimal point source filter is applied to our maps utilising the
information contained within the point source kernel and noise map
realisations.  The filtering techniques used are described in detail
in previous AzTEC publications \citep{scott08,perera08}.  The
resulting AzTEC signal-to-noise maps of Lockman Hole and SXDF are
shown in Fig.~\ref{fig:maps}.  The thick dashed contour of each map
depicts the 50~per cent coverage level, representing a
uniformly-covered region that has a noise level within $\sqrt{2}$ of
that found in the deep central region of that map and beyond which the
survey depth drops off sharply -- a consequence of the particular
observation modes employed.  The AzTEC/SHADES maps are trimmed at this
50~per cent coverage level -- as defined by the co-added weight map
described above -- for all analysis in the following sections.  The
trimmed maps have total sizes of 0.31\,deg$^2$ and 0.37\,deg$^2$ for
LH and SXDF, respectively, and correspond to depths of $0.9 <
\sigma_{\textrm{lh}} < 1.3$\,mJy and $1.0 < \sigma_{\textrm{sxdf}} <
1.7$\,mJy.  The SXDF map is larger and shallower than that of Lockman
Hole due to an observation script error that led to some individual
maps being offset in declination.  We continued to observe the
resulting extended SXDF region after discovery of the error in order
to maximise the usefulness of our entire data set.

The optimal filter is also applied to the co-added point source kernel
and noise maps in order to provide the best model of the point source
response and accurate estimates of the noise properties in our final
maps.  As shown for previously published AzTEC/JCMT maps
\citep{scott08,perera08}, these noise-only maps confirm a highly
Gaussian nature of the underlying noise in the AzTEC/SHADES data 
(see Figure~\ref{fig:hist}).
Accurate representations of the filtered point source kernel and the
noise properties of the maps are critical components of the
simulations described throughout this paper.

\subsection{\label{ssec:astrometry}Astrometry}

We have checked the astrometric accuracy of the AzTEC/SHADES maps 
by stacking (i.e. averaging) the AzTEC flux at the positions of
radio sources in these fields
(techniques described in detail in \citealp{scott08}).
We use a re-reduction of the archival 
VLA 1.4\,GHz continuum data in the Lockman Hole field
\citep{ibar09}
to generate a catalogue of radio sources in the AzTEC/LH field and we
utilise the 100\,$\mu$Jy catalogue of \citet{simpson06} in the
AzTEC/SXDF field.  These catalogues result in stacked detections of
significance 11\,$\sigma$ and 7\,$\sigma$ for AzTEC/LH and AzTEC/SXDF,
respectively.  The stacked data are consistent with no systematic
astrometric offset in either map with the possible exception of a
small offset in declination, $+2.9 \pm 1.3$ arcsec, in the AzTEC/SXDF
field.  Due to the low significance and relatively small size of this
potential offset, no correction is applied to the map.

We can also constrain the random astrometric errors across the AzTEC
maps by measuring the broadening of the stacked signal compared to the
AzTEC point-source response \citep[see][]{scott08}.  This broadening
suggests that the random radial pointing error RMS is
$\sigma_{\textrm{p}} \lsim 4$\,arcsec for both fields.  Broadening of
the stacked signal is also caused by pointing errors in individual
AzTEC observations (final maps are co-additions of dozens of such
observations) and clustering of the radio sources; therefore,
$\sigma_{\textrm{p}} < 4$\,arcsec provides a strong upper limit to the
random astrometric errors in the AzTEC/SHADES maps.

\subsection{\label{ssec:calibration_check}Calibration checks}

Flux density calibration was performed on a nightly basis as described
in Section~\ref{sec:observations}.  All steps in producing the
AzTEC/SHADES maps were performed by the AzTEC instrument team
\citep{wilson08a} and are identical to those employed for other
published AzTEC data sets \citep{scott08,perera08}, thus minimising
systematic differences between these AzTEC surveys.  Flux calibration
is expected to be consistent across all regions of the final
AzTEC/SHADES maps, with each point being sampled numerous times by
each of a large set ($\gg 10$) of observations that span a wide range
of atmospheric conditions and calibrations.

We find no significant systematics between individual observations,
including between the small mosaic and large full-map observations.
Noise properties are consistent across individual observations, with
mapping speeds following the correlation with atmospheric opacity
described in \citet{wilson08a}.  The methods used to remove
atmospheric signal \citep{perera08} produce consistent point source
kernels (shape and amplitude) across all observations,
with the resulting attenuation
of point sources varying about the mean with an RMS of 3.6~per cent
(attenuation corrected for in the final map 
based on the average reduction in flux density). 

We also use the stacking analysis of Section~\ref{ssec:astrometry} as
a check of the relative calibrations between fields based on the
average mm-wave flux of known radio populations.  We find that the
calibrations of the AzTEC/SHADES fields are consistent; the average
1.1--mm fluxes at the location of $S_{1.4\textrm{GHz}} > 100$\,$\mu$Jy
radio sources are $0.59 \pm 0.09$\,mJy and $0.50 \pm 0.07$\,mJy for
the LH and SXDF maps, respectively.  If we remove radio sources that
fall within 9~arcseconds of significant AzTEC `sources' with $S/N \ge
3.5$ or $S/N \le -3.5$ from the stacking analysis, we find that the
stacked average 1.1--mm flux becomes $400 \pm 100\,\mu$Jy and $430 \pm
70\,\mu$Jy for LH and SXDF, respectively.  We also utilise the deeper
radio catalogue available for the LH field to determine that the
average 1.1--mm flux of millimetre-dim $S_{1.4\textrm{GHz}} >
66\,\mu$Jy radio sources is $528 \pm 71\,\mu$Jy, which is consistent
with similar stacks of the AzTEC/COSMOS
\citep[$530\pm87\,\mu$Jy;][]{scott08} and AzTEC/GOODS-N
($439\pm107\,\mu$Jy; using the radio catalogue of \citealp{biggs06}
and the AzTEC map of \citealp{perera08}) surveys.  Assuming the
various fields have similar radio populations, which may not be
strictly true for fields with significant structure
\citep[e.g. AzTEC/COSMOS;][]{austermann08}, these tests show that the
calibration across various AzTEC surveys is consistent within the
measurement errors of the stacks.  Note that since the available radio
catalogue does not cover the entire AzTEC/LH field, we stack only on
radio sources found in deep regions of the radio survey
($\sigma_{1.4\textrm{GHz}} < 13\,\mu$Jy) to ensure a uniform sampling
of $S_{1.4\textrm{GHz}} > 66\,\mu$Jy sources.

\subsection{\label{ssec:catalogs}Catalogues}

Candidate mm-wave sources are identified as local maxima in the
optimally filtered maps that pass a chosen threshold.  Although it is
possible that some local maxima could be due to multiple neighbouring
sources (relative to the FWHM $\approx 18$\,arcsec beamsize) that
blend into one peak, the low $S/N$ of most detections prohibits
deconvolution of potential multi-source peaks.  AzTEC maps are
mean-subtracted (i.e. the background has a zero net contribution) and
SMGs are expected to be sparse, with less than one source per
AzTEC/JCMT beam down to $< 0.1$\,mJy; therefore, source blending is
expected to be rare for bright sources in the AzTEC/SHADES survey,
unless the SMG population is significantly clustered on scales smaller
than 18~arcseconds.  Since very little is known about the SMG
population on these scales, we caution the reader that the following
analysis assumes no clustering at small scales. 
Interferometric
observations over the coming years with the SMA, CARMA, and later with
ALMA will address the clustering question definitively.

The most robust AzTEC/SHADES source candidates are given in
Tables~\ref{table:lh_sources}--\ref{table:sxdf_sources}.  Source
candidates are listed in descending order of detected $S/N$.  Centroid
source positions are determined using flux-squared weighting of the
pixels within 9\,arcsec (FWHM/2) of the local maxima.  The
AzTEC/SHADES survey detects 43 and 21 robust sources with $S/N \ge 4$
in the LH and SXDF fields, respectively.  Additional significant
detections, as defined in Section~\ref{ssec:source_robustness}, are
also listed in the source tables and considered in the analysis of
Section~\ref{sec:counts}.  Multi-wavelength analysis of sources found
in the AzTEC/SHADES survey, including combined
850\,$\micron$/1100\,$\micron$ properties of sources within the
overlapping AzTEC and SCUBA SHADES surveys (Negrello et al., in
prep.), is deferred to future publications.

\subsection{\label{ssec:flux_corrections}Flux corrections}

\begin{figure}
\epsfig{file=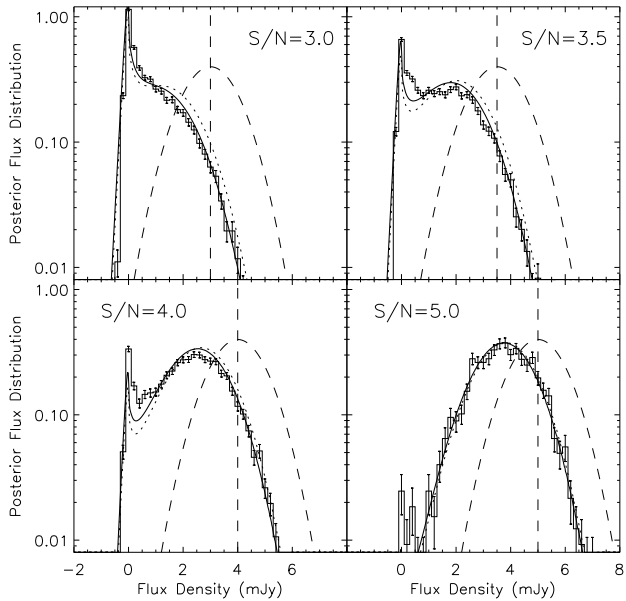, width=1.0\hsize,angle=0}
\caption{Bayesian posterior flux density (PFD; solid curve) compared to the 
intrinsic flux distribution recovered through simulation 
(histogram with 1$\sigma$ Poisson errors of simulation) 
for sources detected at the significance
listed and at a noise level of $\sigma_\textrm{m} = 1.0$\,mJy.  
Also shown are the Bayesian predictions without the correction
for the bias to peak locations in the map (dotted curves).
The dashed vertical line represents the raw measured flux, $S_\textrm{m}$,
while the dashed curve represents the Gaussian probability distribution
that might otherwise be assumed without flux boosting and/or 
false detection considerations. 
Negative flux probabilities are allowable through the 
zero-mean AzTEC point-source kernel and effectively represent the 
probability of a null detection. 
}
\label{fig:deboosted}
\end{figure}

Sources discovered in these blind surveys experience two notable flux biases, both
of which cause the average 
measurement of the flux density of a detected source to be high relative
to its true intrinsic flux, $S_\textrm{i}$.
These flux biases can be very significant ($\sim$~10--50~per cent for sources listed in
Tables~\ref{table:lh_sources}--\ref{table:sxdf_sources}), particularly for the low-significance 
detections that typify SMG surveys.  Therefore, it is important to characterise and correct 
for these biases before fluxes and number counts can be compared to measurements at
other wavelengths.   
Since these biases are a function of survey depth, these corrections are also 
necessary before  
detailed comparisons can be made with other 1.1\,mm surveys.

\begin{table*}
\caption{
AzTEC Lockman Hole (LH)
candidates with $S/N \ge 4$. 
Additional robust AzTEC/LH sources with lower $S/N$ values
are listed in Table~\ref{table:lh_sources_2}.
The columns give: 
1) AzTEC source name, including  RA/DEC centroid position; 
2) Nickname; 
3) Signal-to-noise of the detection in the AzTEC map;
4) Measured 1100--$\micron$ flux density and error; 
5) Flux density and 68~per cent confidence interval, 
as defined in Section~\ref{ssec:source_robustness}, after corrections for
flux boosting and the bias to peak locations in the map;
6) Probability that the source will de-boost
to $S_\textrm{i} < 0$ when assuming the AzTEC/SHADES Bayesian prior.
}
\begin{center}
\begin{tabular}{llcccc}
\hline
\hline
                          &            &       & $S_{1100}$  & $S_{1100}$         &                    \\ 
                          &            &       & (measured)   & (corrected)        &			 \\ 
Source                    & Nickname   & $S/N$ & (mJy)        & (mJy)               & $P(S_{1100}<0)$	 \\ 
\hline
AzTEC\_J105201.98+574049.3 & AzLOCK.1 & 8.2 & $ 7.4\pm0.9$ & $6.6^{+0.9}_{-1.0}$ & 0.000 \\ 
AzTEC\_J105206.08+573622.6 & AzLOCK.2 & 8.2 & $ 7.2\pm0.9$ & $6.4^{+0.9}_{-0.9}$ & 0.000 \\ 
AzTEC\_J105257.18+572105.9 & AzLOCK.3 & 7.5 & $ 7.2\pm1.0$ & $6.2^{+1.1}_{-0.9}$ & 0.000 \\ 
AzTEC\_J105044.47+573318.3 & AzLOCK.4 & 6.7 & $ 6.2\pm0.9$ & $5.3^{+0.9}_{-1.0}$ & 0.000 \\ 
AzTEC\_J105403.76+572553.7 & AzLOCK.5 & 6.4 & $ 5.9\pm0.9$ & $4.9^{+1.0}_{-0.9}$ & 0.000 \\ 
AzTEC\_J105241.89+573551.7 & AzLOCK.6 & 6.2 & $ 5.6\pm0.9$ & $4.8^{+0.8}_{-1.0}$ & 0.000 \\ 
AzTEC\_J105203.89+572700.5 & AzLOCK.7 & 6.0 & $ 5.7\pm0.9$ & $4.8^{+0.9}_{-1.1}$ & 0.000 \\ 
AzTEC\_J105201.14+572443.0 & AzLOCK.8 & 6.0 & $ 5.6\pm0.9$ & $4.7^{+1.0}_{-1.0}$ & 0.000 \\ 
AzTEC\_J105214.22+573327.4 & AzLOCK.9 & 5.7 & $ 5.0\pm0.9$ & $4.1^{+0.9}_{-0.9}$ & 0.000 \\ 
AzTEC\_J105406.44+573309.6 & AzLOCK.10 & 5.5 & $ 5.1\pm0.9$ & $4.1^{+0.9}_{-1.0}$ & 0.000 \\ 
AzTEC\_J105130.29+573807.2 & AzLOCK.11 & 5.4 & $ 4.8\pm0.9$ & $3.8^{+1.0}_{-0.9}$ & 0.000 \\ 
AzTEC\_J105217.23+573501.4 & AzLOCK.12 & 5.4 & $ 4.7\pm0.9$ & $3.8^{+0.9}_{-1.0}$ & 0.000 \\ 
AzTEC\_J105140.64+574324.6 & AzLOCK.13 & 5.3 & $ 5.3\pm1.0$ & $4.1^{+1.0}_{-1.1}$ & 0.000 \\ 
AzTEC\_J105220.24+573955.1 & AzLOCK.14 & 5.2 & $ 4.6\pm0.9$ & $3.6^{+0.9}_{-0.9}$ & 0.000 \\ 
AzTEC\_J105256.32+574227.5 & AzLOCK.15 & 5.2 & $ 4.8\pm0.9$ & $3.7^{+0.9}_{-1.0}$ & 0.000 \\ 
AzTEC\_J105341.50+573215.9 & AzLOCK.16 & 5.2 & $ 4.7\pm0.9$ & $3.7^{+1.0}_{-1.0}$ & 0.000 \\ 
AzTEC\_J105319.47+572105.3 & AzLOCK.17 & 5.0 & $ 4.7\pm0.9$ & $3.6^{+1.0}_{-1.0}$ & 0.001 \\ 
AzTEC\_J105225.16+573836.7 & AzLOCK.18 & 4.8 & $ 4.2\pm0.9$ & $3.2^{+1.0}_{-0.9}$ & 0.001 \\ 
AzTEC\_J105129.55+573649.2 & AzLOCK.19 & 4.8 & $ 4.3\pm0.9$ & $3.2^{+1.0}_{-0.9}$ & 0.002 \\ 
AzTEC\_J105345.53+571647.0 & AzLOCK.20 & 4.7 & $ 4.7\pm1.0$ & $3.4^{+1.1}_{-1.1}$ & 0.004 \\ 
AzTEC\_J105131.41+573134.1 & AzLOCK.21 & 4.7 & $ 4.1\pm0.9$ & $3.1^{+0.9}_{-1.0}$ & 0.003 \\ 
AzTEC\_J105256.49+572356.7 & AzLOCK.22 & 4.7 & $ 4.5\pm1.0$ & $3.2^{+1.1}_{-1.0}$ & 0.004 \\ 
AzTEC\_J105321.96+571717.8 & AzLOCK.23 & 4.5 & $ 4.4\pm1.0$ & $3.1^{+1.0}_{-1.1}$ & 0.007 \\ 
AzTEC\_J105238.46+572436.8 & AzLOCK.24 & 4.5 & $ 4.3\pm0.9$ & $3.0^{+1.0}_{-1.1}$ & 0.007 \\ 
AzTEC\_J105107.06+573442.2 & AzLOCK.25 & 4.4 & $ 3.9\pm0.9$ & $2.7^{+1.0}_{-0.9}$ & 0.008 \\ 
AzTEC\_J105059.75+571636.7 & AzLOCK.26 & 4.3 & $ 4.6\pm1.0$ & $3.1^{+1.1}_{-1.3}$ & 0.016 \\ 
AzTEC\_J105218.64+571852.9 & AzLOCK.27 & 4.3 & $ 4.3\pm1.0$ & $2.9^{+1.1}_{-1.1}$ & 0.013 \\ 
AzTEC\_J105045.11+573650.4 & AzLOCK.28 & 4.3 & $ 4.1\pm1.0$ & $2.7^{+1.1}_{-1.1}$ & 0.015 \\ 
AzTEC\_J105123.33+572200.8 & AzLOCK.29 & 4.2 & $ 4.0\pm0.9$ & $2.7^{+1.1}_{-1.1}$ & 0.016 \\ 
AzTEC\_J105238.09+573003.4 & AzLOCK.30 & 4.2 & $ 3.8\pm0.9$ & $2.6^{+0.9}_{-1.1}$ & 0.014 \\ 
AzTEC\_J105425.31+573707.8 & AzLOCK.31 & 4.2 & $ 5.2\pm1.2$ & $3.1^{+1.4}_{-1.6}$ & 0.038 \\ 
AzTEC\_J105041.16+572129.6 & AzLOCK.32 & 4.2 & $ 4.1\pm1.0$ & $2.7^{+1.1}_{-1.2}$ & 0.020 \\ 
AzTEC\_J105245.93+573121.2 & AzLOCK.33 & 4.2 & $ 3.8\pm0.9$ & $2.5^{+1.0}_{-1.0}$ & 0.016 \\ 
AzTEC\_J105238.35+572324.4 & AzLOCK.34 & 4.1 & $ 4.0\pm1.0$ & $2.6^{+1.0}_{-1.2}$ & 0.023 \\ 
AzTEC\_J105355.84+572954.7 & AzLOCK.35 & 4.1 & $ 3.7\pm0.9$ & $2.5^{+1.0}_{-1.1}$ & 0.021 \\ 
AzTEC\_J105349.58+571604.3 & AzLOCK.36 & 4.1 & $ 4.3\pm1.1$ & $2.7^{+1.2}_{-1.3}$ & 0.032 \\ 
AzTEC\_J105152.72+571334.5 & AzLOCK.37 & 4.1 & $ 4.1\pm1.0$ & $2.6^{+1.1}_{-1.3}$ & 0.033 \\ 
AzTEC\_J105116.44+573209.9 & AzLOCK.38 & 4.0 & $ 3.5\pm0.9$ & $2.4^{+1.0}_{-1.1}$ & 0.025 \\ 
AzTEC\_J105212.26+571552.5 & AzLOCK.39 & 4.0 & $ 4.0\pm1.0$ & $2.5^{+1.1}_{-1.3}$ & 0.035 \\ 
AzTEC\_J105226.58+573355.0 & AzLOCK.40 & 4.0 & $ 3.5\pm0.9$ & $2.3^{+1.0}_{-1.0}$ & 0.026 \\ 
AzTEC\_J105116.34+574027.3 & AzLOCK.41 & 4.0 & $ 3.7\pm0.9$ & $2.4^{+1.1}_{-1.2}$ & 0.036 \\ 
AzTEC\_J105058.27+571842.8 & AzLOCK.42 & 4.0 & $ 3.9\pm1.0$ & $2.4^{+1.2}_{-1.2}$ & 0.042 \\ 
AzTEC\_J105153.10+572122.7 & AzLOCK.43 & 4.0 & $ 3.8\pm1.0$ & $2.4^{+1.1}_{-1.2}$ & 0.039 \\ 
\hline
\end{tabular}
\end{center}
\label{table:lh_sources}
\end{table*}

\newpage

\begin{table*}
\caption{AzTEC/LH source
candidates with $S/N < 4$.
Columns as described in Table~\ref{table:lh_sources}.
}
\begin{center}
\begin{tabular}{llcccc}
\hline
\hline
                          &            &       & $S_{1100}$ & $S_{1100}$         &                     \\ 
                          &            &       & (measured)   & (corrected)        &			  \\
Source                    & Nickname   & $S/N$ & (mJy)        & (mJy)               & $P(S_{1100}<0)$	  \\ 
\hline
AzTEC\_J105241.87+573406.1 & AzLOCK.44 & 3.9 & $ 3.5\pm0.9$ & $2.3^{+1.0}_{-1.1}$ & 0.033 \\ 
AzTEC\_J105154.82+573824.6 & AzLOCK.45 & 3.9 & $ 3.5\pm0.9$ & $2.2^{+1.0}_{-1.1}$ & 0.035 \\ 
AzTEC\_J105210.75+571433.8 & AzLOCK.46 & 3.9 & $ 3.9\pm1.0$ & $2.4^{+1.2}_{-1.3}$ & 0.047 \\ 
AzTEC\_J105306.80+573032.7 & AzLOCK.47 & 3.9 & $ 3.6\pm0.9$ & $2.3^{+1.0}_{-1.2}$ & 0.037 \\ 
AzTEC\_J105431.31+572543.3 & AzLOCK.48 & 3.9 & $ 4.0\pm1.0$ & $2.4^{+1.3}_{-1.3}$ & 0.052 \\ 
AzTEC\_J105340.49+572755.0 & AzLOCK.49 & 3.9 & $ 3.6\pm0.9$ & $2.2^{+1.1}_{-1.1}$ & 0.039 \\ 
AzTEC\_J105205.59+572916.1 & AzLOCK.50 & 3.9 & $ 3.5\pm0.9$ & $2.2^{+1.0}_{-1.2}$ & 0.042 \\ 
AzTEC\_J105035.90+573332.1 & AzLOCK.51 & 3.9 & $ 3.7\pm1.0$ & $2.2^{+1.2}_{-1.2}$ & 0.050 \\ 
AzTEC\_J105206.79+574537.5 & AzLOCK.52 & 3.9 & $ 4.2\pm1.1$ & $2.4^{+1.3}_{-1.6}$ & 0.070 \\ 
AzTEC\_J105435.20+572715.9 & AzLOCK.53 & 3.9 & $ 4.0\pm1.0$ & $2.4^{+1.2}_{-1.5}$ & 0.062 \\ 
AzTEC\_J105351.57+572648.8 & AzLOCK.54 & 3.8 & $ 3.5\pm0.9$ & $2.1^{+1.1}_{-1.2}$ & 0.050 \\ 
AzTEC\_J105153.94+571034.3 & AzLOCK.55 & 3.8 & $ 4.6\pm1.2$ & $2.4^{+1.3}_{-2.0}$ & 0.094 \\ 
AzTEC\_J105203.84+572522.7 & AzLOCK.56 & 3.8 & $ 3.6\pm0.9$ & $2.1^{+1.1}_{-1.3}$ & 0.055 \\ 
AzTEC\_J105251.38+572609.9 & AzLOCK.57 & 3.8 & $ 3.6\pm0.9$ & $2.1^{+1.2}_{-1.2}$ & 0.056 \\ 
AzTEC\_J105243.78+574042.6 & AzLOCK.58 & 3.8 & $ 3.4\pm0.9$ & $2.1^{+1.1}_{-1.2}$ & 0.053 \\ 
AzTEC\_J105044.92+573030.0 & AzLOCK.59 & 3.8 & $ 3.4\pm0.9$ & $2.1^{+1.1}_{-1.2}$ & 0.054 \\ 
AzTEC\_J105345.63+572645.8 & AzLOCK.60 & 3.8 & $ 3.5\pm0.9$ & $2.1^{+1.1}_{-1.2}$ & 0.056 \\ 
AzTEC\_J105257.19+572248.5 & AzLOCK.61 & 3.8 & $ 3.7\pm1.0$ & $2.1^{+1.2}_{-1.3}$ & 0.063 \\ 
AzTEC\_J105211.61+573510.7 & AzLOCK.62 & 3.8 & $ 3.3\pm0.9$ & $2.0^{+1.0}_{-1.2}$ & 0.056 \\ 
AzTEC\_J105406.14+572042.0 & AzLOCK.63 & 3.7 & $ 3.7\pm1.0$ & $2.1^{+1.2}_{-1.4}$ & 0.074 \\ 
AzTEC\_J105310.94+573435.6 & AzLOCK.64 & 3.7 & $ 3.3\pm0.9$ & $2.0^{+1.1}_{-1.2}$ & 0.059 \\ 
AzTEC\_J105258.39+573935.4 & AzLOCK.65 & 3.7 & $ 3.4\pm0.9$ & $2.0^{+1.1}_{-1.2}$ & 0.061 \\ 
AzTEC\_J105351.46+573058.2 & AzLOCK.66 & 3.7 & $ 3.4\pm0.9$ & $2.0^{+1.1}_{-1.2}$ & 0.064 \\ 
AzTEC\_J105045.33+572924.4 & AzLOCK.67 & 3.7 & $ 3.4\pm0.9$ & $1.9^{+1.1}_{-1.2}$ & 0.065 \\ 
AzTEC\_J105325.86+572247.3 & AzLOCK.68 & 3.7 & $ 3.5\pm0.9$ & $2.0^{+1.1}_{-1.3}$ & 0.071 \\ 
AzTEC\_J105059.74+573245.6 & AzLOCK.69 & 3.7 & $ 3.3\pm0.9$ & $1.9^{+1.1}_{-1.2}$ & 0.064 \\ 
AzTEC\_J105121.65+573333.6 & AzLOCK.70 & 3.7 & $ 3.2\pm0.9$ & $1.9^{+1.0}_{-1.2}$ & 0.064 \\ 
AzTEC\_J105407.02+572957.7 & AzLOCK.71 & 3.7 & $ 3.4\pm0.9$ & $1.9^{+1.1}_{-1.3}$ & 0.071 \\ 
AzTEC\_J105132.73+574022.1 & AzLOCK.72 & 3.7 & $ 3.4\pm0.9$ & $1.9^{+1.1}_{-1.2}$ & 0.069 \\ 
AzTEC\_J105157.08+574057.6 & AzLOCK.73 & 3.7 & $ 3.3\pm0.9$ & $1.9^{+1.1}_{-1.2}$ & 0.068 \\ 
AzTEC\_J105246.38+571742.5 & AzLOCK.74 & 3.7 & $ 3.6\pm1.0$ & $1.9^{+1.2}_{-1.4}$ & 0.087 \\ 
AzTEC\_J105309.72+571700.1 & AzLOCK.75 & 3.7 & $ 3.6\pm1.0$ & $1.9^{+1.2}_{-1.4}$ & 0.087 \\ 
AzTEC\_J105228.45+573258.0 & AzLOCK.76 & 3.7 & $ 3.2\pm0.9$ & $1.9^{+1.0}_{-1.2}$ & 0.067 \\ 
AzTEC\_J105148.13+574122.5 & AzLOCK.77 & 3.7 & $ 3.3\pm0.9$ & $1.9^{+1.1}_{-1.3}$ & 0.073 \\ 
AzTEC\_J105349.75+573352.4 & AzLOCK.78 & 3.7 & $ 3.4\pm0.9$ & $1.9^{+1.1}_{-1.3}$ & 0.076 \\ 
AzTEC\_J105232.60+571540.3 & AzLOCK.79 & 3.7 & $ 3.6\pm1.0$ & $1.9^{+1.2}_{-1.5}$ & 0.088 \\ 
AzTEC\_J105418.55+573447.5 & AzLOCK.80 & 3.7 & $ 3.7\pm1.0$ & $1.9^{+1.1}_{-1.5}$ & 0.093 \\ 
AzTEC\_J105321.70+572308.3 & AzLOCK.81 & 3.6 & $ 3.4\pm0.9$ & $1.9^{+1.1}_{-1.4}$ & 0.083 \\ 
AzTEC\_J105136.91+573758.1 & AzLOCK.82 & 3.6 & $ 3.2\pm0.9$ & $1.8^{+1.0}_{-1.3}$ & 0.079 \\ 
AzTEC\_J105343.81+572543.6 & AzLOCK.83 & 3.6 & $ 3.3\pm0.9$ & $1.8^{+1.0}_{-1.4}$ & 0.090 \\ 
AzTEC\_J105230.53+572210.0 & AzLOCK.84 & 3.6 & $ 3.4\pm1.0$ & $1.8^{+1.0}_{-1.6}$ & 0.099 \\ 
AzTEC\_J105036.93+573228.9 & AzLOCK.85 & 3.6 & $ 3.3\pm0.9$ & $1.8^{+1.0}_{-1.5}$ & 0.096 \\ 
AzTEC\_J105037.18+572844.9 & AzLOCK.86 & 3.6 & $ 3.3\pm0.9$ & $1.7^{+0.9}_{-1.5}$ & 0.099 \\ 
\hline
\end{tabular}
\end{center}
\label{table:lh_sources_2}
\end{table*}

\newpage

\begin{table*}
\caption{AzTEC/SXDF source
candidates. 
Columns as described in Table~\ref{table:lh_sources}.
}
\begin{center}
\begin{tabular}{llcccc}
\hline
\hline
                          &            &       & $S_{1100}$  & $S_{1100}$         &                     \\ 
                          &            &       & (measured)   & (corrected)        &			  \\ 
Source                    & Nickname   & $S/N$ & (mJy)        & (mJy)               & $P(S_{1100}<0)$	  \\ 
\hline
AzTEC\_J021738.52$-$043330.3 & AzSXDF.1 & 5.2 & $ 7.4\pm1.4$ & $5.3^{+1.4}_{-1.7}$ & 0.002 \\ 
AzTEC\_J021745.76$-$044747.8 & AzSXDF.2 & 4.8 & $ 5.4\pm1.1$ & $4.0^{+1.1}_{-1.3}$ & 0.003 \\ 
AzTEC\_J021754.97$-$044723.9 & AzSXDF.3 & 4.8 & $ 5.3\pm1.1$ & $3.8^{+1.2}_{-1.2}$ & 0.004 \\ 
AzTEC\_J021831.27$-$043911.9 & AzSXDF.4 & 4.8 & $ 6.9\pm1.4$ & $4.4^{+1.7}_{-1.6}$ & 0.011 \\ 
AzTEC\_J021742.10$-$045626.7 & AzSXDF.5 & 4.7 & $ 5.1\pm1.1$ & $3.6^{+1.2}_{-1.2}$ & 0.005 \\ 
AzTEC\_J021842.39$-$045932.7 & AzSXDF.6 & 4.6 & $ 5.8\pm1.3$ & $4.0^{+1.3}_{-1.6}$ & 0.010 \\ 
AzTEC\_J021655.80$-$044532.2 & AzSXDF.7 & 4.6 & $ 6.3\pm1.4$ & $4.0^{+1.6}_{-1.7}$ & 0.019 \\ 
AzTEC\_J021742.13$-$043135.6 & AzSXDF.8 & 4.5 & $ 6.4\pm1.4$ & $4.0^{+1.6}_{-1.8}$ & 0.025 \\ 
AzTEC\_J021823.10$-$051136.7 & AzSXDF.9 & 4.3 & $ 6.9\pm1.6$ & $3.8^{+1.9}_{-2.4}$ & 0.061 \\ 
AzTEC\_J021816.07$-$045512.2 & AzSXDF.10 & 4.3 & $ 4.7\pm1.1$ & $3.1^{+1.2}_{-1.3}$ & 0.018 \\ 
AzTEC\_J021708.04$-$045615.3 & AzSXDF.11 & 4.3 & $ 5.5\pm1.3$ & $3.3^{+1.5}_{-1.7}$ & 0.039 \\ 
AzTEC\_J021708.03$-$044256.8 & AzSXDF.12 & 4.2 & $ 5.9\pm1.4$ & $3.3^{+1.7}_{-1.9}$ & 0.053 \\ 
AzTEC\_J021829.13$-$045448.2 & AzSXDF.13 & 4.2 & $ 4.4\pm1.1$ & $2.8^{+1.2}_{-1.4}$ & 0.028 \\ 
AzTEC\_J021740.55$-$044609.1 & AzSXDF.14 & 4.1 & $ 4.8\pm1.2$ & $2.9^{+1.3}_{-1.5}$ & 0.037 \\ 
AzTEC\_J021754.76$-$044417.5 & AzSXDF.15 & 4.1 & $ 4.8\pm1.2$ & $2.9^{+1.3}_{-1.5}$ & 0.037 \\ 
AzTEC\_J021716.24$-$045808.4 & AzSXDF.16 & 4.1 & $ 5.0\pm1.2$ & $2.9^{+1.5}_{-1.6}$ & 0.044 \\ 
AzTEC\_J021711.62$-$044315.1 & AzSXDF.17 & 4.1 & $ 5.6\pm1.4$ & $3.1^{+1.6}_{-2.0}$ & 0.064 \\ 
AzTEC\_J021724.48$-$043144.5 & AzSXDF.18 & 4.1 & $ 6.1\pm1.5$ & $3.1^{+1.5}_{-2.6}$ & 0.091 \\ 
AzTEC\_J021906.24$-$045333.4 & AzSXDF.19 & 4.0 & $ 6.5\pm1.6$ & $3.3^{+0.9}_{-3.3}$ & 0.118$^a$ \\ 
AzTEC\_J021742.13$-$050723.4 & AzSXDF.20 & 4.0 & $ 5.7\pm1.4$ & $2.9^{+1.3}_{-2.6}$ & 0.096 \\ 
AzTEC\_J021809.81$-$050444.8 & AzSXDF.21 & 4.0 & $ 5.0\pm1.3$ & $2.6^{+1.6}_{-1.8}$ & 0.070 \\ 
AzTEC\_J021827.89$-$045320.5 & AzSXDF.22 & 3.9 & $ 4.2\pm1.1$ & $2.5^{+1.2}_{-1.5}$ & 0.057 \\ 
AzTEC\_J021820.23$-$045738.7 & AzSXDF.23 & 3.9 & $ 4.3\pm1.1$ & $2.5^{+1.3}_{-1.6}$ & 0.060 \\ 
AzTEC\_J021832.33$-$045632.7 & AzSXDF.24 & 3.8 & $ 4.1\pm1.1$ & $2.3^{+1.3}_{-1.5}$ & 0.065 \\ 
AzTEC\_J021802.42$-$050018.4 & AzSXDF.25 & 3.8 & $ 4.2\pm1.1$ & $2.3^{+1.2}_{-1.7}$ & 0.081 \\ 
AzTEC\_J021756.39$-$045242.5 & AzSXDF.26 & 3.8 & $ 4.0\pm1.1$ & $2.1^{+1.3}_{-1.5}$ & 0.076 \\ 
AzTEC\_J021741.50$-$050218.0 & AzSXDF.27 & 3.8 & $ 4.4\pm1.2$ & $2.3^{+1.1}_{-2.0}$ & 0.096 \\ 
AzTEC\_J021806.97$-$044941.9 & AzSXDF.28 & 3.7 & $ 3.9\pm1.1$ & $2.0^{+1.1}_{-1.7}$ & 0.091 \\ 
\hline
\end{tabular}
\end{center}
Notes:
$a)$ Source is included in order to have a complete list of candidates with $S/N$~$\geq$~4, 
despite its relatively high null probability.
\label{table:sxdf_sources}
\end{table*}

The primary flux bias in SMG surveys is commonly referred to as `flux boosting'
and is due to the combination of a source density that 
increases sharply with decreasing flux
and the blind 
nature of 
the survey 
(i.e. sources have previously unknown positions); 
see \citet{hogg98} for a full description of this effect.  
We employ an advanced version of the Bayesian methods 
of \citet{coppin05,coppin06} to correct 
for flux boosting and generate a full posterior flux density (PFD)
probability distribution for each source candidate.  
The Bayesian approach requires a prior in the form of the 
assumed number density of sources projected on the sky 
(i.e. `number counts') as a function of flux.
We use the iterative method of \citet{austermann08} to determine the 
most appropriate prior.  
We begin by using the SCUBA/SHADES \citep{coppin06} 850--$\micron$
number counts, scaled to 1.1\,mm through an initial assumption of the
850/1100\,$\micron$ flux ratio,
as the initial prior.  
The prior is then iteratively adjusted using the empirical number counts 
of this survey (Section~\ref{ssec:nc_algorithm}), which quickly converges
within a few iterations.  
As the widest-area deep millimetre survey to date, these
iterative AzTEC/SHADES results
provide the best 1.1--mm blank-field source number density 
prior available.  

A second notable flux bias results from sources being defined as 
local maxima in the map.  
Since the position of the source is not independently known,
nearby positive noise inevitably induces positional errors
and this noise can combine with the off-centre
beam-convolved flux of the source to outshine 
the true source being measured,
thus resulting in an average positive flux bias in the local maximum 
that is taken as the measurement.     
The bias is independent of the aforementioned `flux boosting' 
(the Bayesian prior is a noiseless calculation) and is instead a
systematic of the actual measurement, as opposed to an effect of
the luminosity function being surveyed.  
This bias to peaks (or `noise gradient bias', e.g. \citealp{ivison07})
is minimised by optimally filtering the map 
for point sources, but can 
still be a significant factor for low-significance sources.

We characterise and quantify the bias to peak locations through 
10,000 simulations of the LH and SXDF maps.  
These simulated maps are generated by populating the noise-only maps
with the flux-scaled point source kernel at random locations
drawn from a uniform distribution and in 
accordance with a number counts distribution 
that is consistent with the final AzTEC/SHADES counts
(Section~\ref{ssec:nc_algorithm}).   
We generate simulated PFDs by cataloguing the input flux ($S_\textrm{i}$) associated 
with each source measurement ($S_\textrm{m}$,$\sigma_\textrm{m}$)
recovered in the simulated maps.
These simulated PFDs are compared to the Bayesian estimate to 
characterise the remaining bias (e.g. Fig.~\ref{fig:deboosted}), 
which comes primarily from the bias to peak locations.  
Through comparison of the PFDs over  
the flux range under investigation here
($S_\textrm{i} >$ 1\,mJy)
and for detections with $S/N \ge$ 3 
, we find that the average 
flux bias incurred for an AzTEC/JCMT measurement of ($S_\textrm{m}$,$\sigma_\textrm{m}$) is 
well described by the equation
\be
\label{eq:biaspeak}
{b_\mathrm{peak}(S_\textrm{m},\sigma_\textrm{m})} = {{\alpha \sigma_\textrm{m}} \over \sqrt{2 \pi}} \exp{\left({-{\beta^2 S_\textrm{m}^2}} \over {2 \sigma_\textrm{m}^2}\right)}
\ee
with $\alpha=1$ and $\beta=0.4$.     
In this form, the bias is modelled as $\alpha$ effective independent 
noise elements that lie 
at a radial distance from the true source that is equivalent to 
that where the fractional flux of the Gaussian beam (relative to maximum) is $\beta$. 
Although this bias is relatively small in flux, 
it can have a strong effect on the Bayesian probability 
densities of low $S/N$ detections
(e.g. 50~per cent overestimate in probability 
that $S_\textrm{i} = S_\textrm{m}$ for a $S_\textrm{m} = 3 \pm 1$\,mJy measurement;
see Fig.~\ref{fig:deboosted}).
We note that the estimates provided by \Eq\ref{eq:biaspeak} are 
significantly smaller than the generalised case provided by 
Equation~B19 of \citet{ivison07} and are specifically tailored 
to AzTEC/JCMT scanning observations through simulation.  
Maps with a significantly different response to 
point sources (e.g. different beamsize or mapping strategy) may require
a reevaluation of the functional form and parameter values 
of \Eq\ref{eq:biaspeak}.  
 
We correct for the bias to peak locations by subtracting $b_\textrm{peak}$ 
from the measured flux, $S_\textrm{m}$, before calculating the Bayesian estimated 
PFD.  
The differences between the Bayesian PFDs with (solid curves) and 
without (dashed curves) this secondary bias correction can be 
seen in Fig.~\ref{fig:deboosted} for several $S/N$ detection levels.
Analysis of past surveys typically ignored this bias, but largely avoided
its effects by restricting the analysis to only the most 
significant sources (e.g. $S/N$ $\gsim$ 4).  

The de-boosted flux value listed for each source
in Tables~\ref{table:lh_sources}--\ref{table:sxdf_sources}
is taken as the flux  
at the PFD local maximum nearest the detected flux, $S_\textrm{m}$
(see Fig.~\ref{fig:deboosted}).
Our improved estimate of the significant biases at work 
for low significance detections leads to accurate PFDs 
down to at least $S/N =3$, thus allowing us to utilise 
more of the maps' information when 
conducting the source-list driven number counts analysis 
of Section~\ref{sec:counts}.

\subsection{\label{ssec:source_robustness}Source robustness and false detections}

\begin{figure}
\epsfig{file=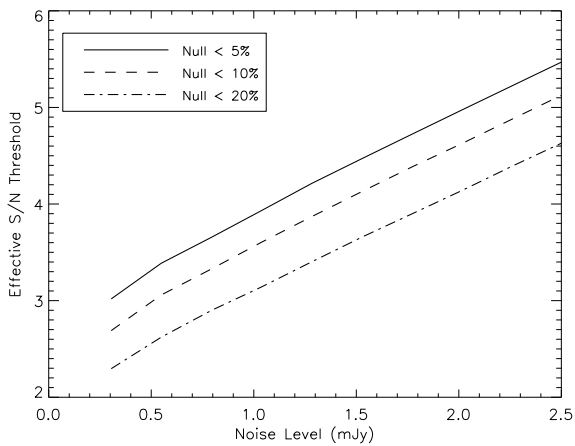, width=1.0\hsize,angle=0}
\caption{Effective $S/N$ threshold as a function of 
noise level for the given null-threshold values 
when assuming the AzTEC/SHADES number counts as the 
Bayesian prior.  
These curves represent constant levels of robustness
for an SMG detection using AzTEC/JCMT data.
These values are unique to the particular Bayesian prior 
(number counts) assumed.  
}
\label{fig:eff_sn}
\end{figure}

The effects of flux boosting (Section~\ref{ssec:flux_corrections})
make $S/N$ a less than ideal measure of source robustness; therefore, 
we include an estimate of the total probability that the source will 
de-boost to $S_\textrm{i} < 0$, $P(S_{1100}<0)$, 
as determined from the integrated Bayesian PFD, 
in the source lists of Tables~\ref{table:lh_sources}--\ref{table:sxdf_sources}.  
This provides a better metric than just $S/N$ for the \textit{relative} robustness 
of the source detections, due to its dependence on both 
$S_\textrm{m}$ and $\sigma_\textrm{m}$, rather than just the ratio $S_\textrm{m}/\sigma_\textrm{m}$
\citep[see also][]{coppin06}. 
We have restricted Tables~\ref{table:lh_sources}--\ref{table:sxdf_sources}
to include only the most robust AzTEC/SHADES 
sources with $P(S_{1100}<0) \le 0.1$.
The effective $S/N$, as a function of $\sigma_\textrm{m}$, of this
10~per cent `null-threshold' is plotted in Fig.~\ref{fig:eff_sn}.
We note that the \textit{absolute} value of $P(S_{1100}<0)$ is highly 
sensitive to the Bayesian prior used. 
For example, if we instead assumed the results of the 
relatively source-rich AzTEC/GOODS-N survey
\citep{perera08} as the Bayesian prior, the number of sources 
in AzTEC/LH passing the 10~per cent null-threshold increases from 86 
to 221.
Therefore, it is important to consider the 
priors used when comparing the number of `detections' 
in various surveys of this type.    
However, we note that the effect of the choice of prior 
on the resulting number counts (Section~\ref{sec:counts})
is much less substantial, as the apparent change in 
the number of `detections' is largely counteracted by the 
survey completeness corresponding to the particular prior used.  

\begin{figure}
\epsfig{file=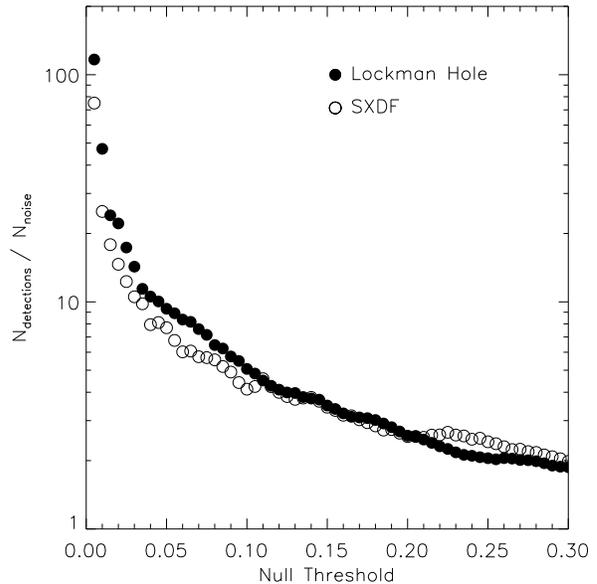, width=1.0\hsize,angle=0}
\caption{
Ratio of the total number of detections to the number 
of significant noise-only peaks as a function 
of null-threshold for the 50~per cent coverage region
of AzTEC/LH 
and
AzTEC/SXDF.
This provides an estimate of the relative number of false detections
expected beyond a given source threshold, i.e. below a 
given null-threshold.  
}
\label{fig:fdr}
\end{figure}

The number of false detections in a given source catalogue depends
strongly on the chosen threshold for what is, and what is not, defined
as a source.  Due to the relatively large 18\,arcsecond beamsize of
AzTEC on the JCMT, the AzTEC/SHADES maps are expected to become
significantly `full' of sources (on average one source per beam) when
considering the expected high density of sources with 1.1--mm fluxes
$< 0.1$\,mJy.  Various estimates of the false detection rate of
AzTEC/JCMT maps are explored in \citet{perera08}, who conclude that
the average number of significant noise peaks in the jack-knifed
noise-only maps provide a conservative overestimate of the number of
false detections in the map 
(a consequence of true source signal adding both positive and negative fluxes 
to the underlying noise distribution to our zero-mean maps).  
The ratio of number of sources in the signal map to
the average number found in the corresponding noise-only maps is
plotted as a function of null-threshold in Fig.~\ref{fig:fdr} for the
LH and SXDF fields.


\section{SMG Number Counts}
\label{sec:counts}

In this section, we present the sky-projected densities of 1.1--mm sources
in the AzTEC/SHADES survey and the methods by which they are determined.
These methods represent an extended and improved version of the algorithms 
outlined in the SCUBA/SHADES number counts paper \citep{coppin06}.
In Section~\ref{ssec:nc_fits} we provide parametric fits to the number count results
of the combined surveys.  
These number count results provide a useful measure of the SMG 
population through which 
we compare those found
in other fields,
at different wavelengths, and that predicted by various models 
and simulations (Section~\ref{sec:discussion}).

\subsection{\label{ssec:nc_algorithm}Number counts: algorithm and results}

We calculate source number counts using the bootstrap sampling methods outlined in
\citet{austermann08}, which are motivated by those
used to determine the SCUBA/SHADES number counts \citep{coppin06}.
In this method, the catalogues of continuous source PFDs are sampled
at random and with replacement \citep[e.g.][]{press92} 
in order to determine specific intrinsic fluxes for the sources in the catalogue.
These samples are binned to produce both differential and integral source
counts as a function of intrinsic flux.
This sampling process is repeated 100,000 times to  
provide sufficient sampling of the source count probability distribution. 
Sampling variance is injected by Poisson deviating the 
number of sources sampled in each of the 100,000 iterations 
around the actual number of sources
detected in the map.
Number count results are taken as the mean of each bin
and the distribution across the iterations 
is used to characterise the associated uncertainty.
The counts are then corrected for completeness,
using estimates derived from simulation, and scaled for survey area.  
The resulting number counts are then taken as the new Bayesian prior
and the entire process, including producing new catalogues 
of sources and their PFDs, is repeated in the iterative-prior process
described in Section~\ref{ssec:flux_corrections}. 
For each iteration of the prior, 
the number counts are calculated for both the LH and SXDF
surveys independently and also for the two surveys combined.
The Bayesian prior chosen for the next iteration is always
taken as the best fit to the combined result.
This iterative procedure minimises our bias to the number counts 
assumed in the Bayesian calculations.

Previous surveys using a similar bootstrapping technique 
\citep{coppin06,perera08,austermann08}
limited the source catalogue to those sources with 
negative flux probability of $P(S_\textrm{i} < 0) < 0.05$, i.e. a null-threshold of 5~per cent.
This null-threshold value was historically used to limit the number of false detections
to a near negligible amount and to render the bias to peak locations 
relatively insignificant.  
However, the false detection probability is inherently accounted for
in the bootstrap sampling method if accurate PFDs are used. 
As discussed in Section~\ref{sec:maps}, our bias corrections result in PFDs 
that are accurate for all source candidates with $S/N \ge 3$, 
and possibly lower significance (currently untested below $S/N = 3$). 
The PFDs are particularly accurate in the 
$S_\textrm{i} \ge 1$\,mJy flux range 
considered in this analysis.  
Since the traditional null-threshold of 5~per cent 
would limit the AzTEC/SHADES source candidate list to just 
those with $S/N \gsim 4$ (Fig.~\ref{fig:eff_sn}), 
we explore fainter sources in the data set 
with the use of higher null-thresholds that incorporate a larger 
catalogue of source candidates in the derivation of source count densities. 

We derive combined AzTEC/SHADES number counts using 
null-thresholds of 5, 10, and 20~per cent. 
The 20~per cent threshold represents the lowest significance 
tested in our simulations 
(effective $S/N \gsim 3$) and safely avoids complications
related to source confusion by keeping the density 
of detections sufficiently low.
The results are consistent for all three 
threshold values tested and the 
variations between the results are, in general, 
much smaller than the formal 68~per cent uncertainty 
of the number count estimates.
We have verified through simulation that 
the use of the higher null-threshold values
supplies additional data without 
introducing any significant biases 
or systematics (Section~\ref{ssec:nc_simulations}).

\begin{figure}
\epsfig{file=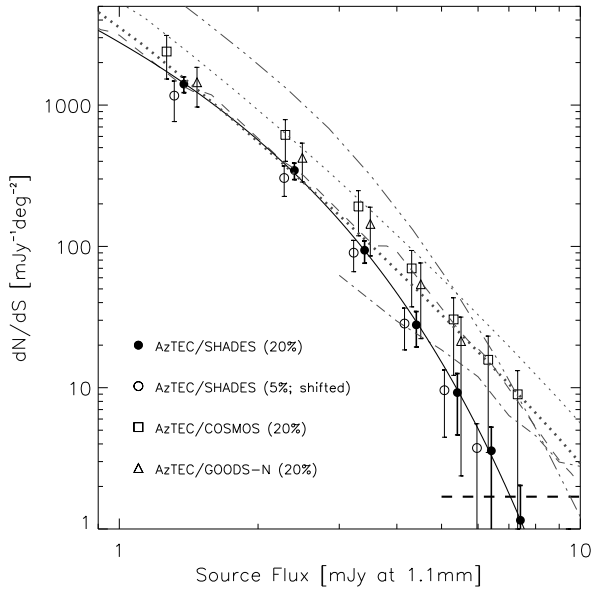, width=1.0\hsize,angle=0}
\caption{Differential number counts for the combined
AzTEC/SHADES survey in 1\,mJy wide bins. 
The per cent value in parentheses represents the null-threshold
used for each data set.
The 5~per cent threshold results of AzTEC/SHADES 
have been artificially displaced
to the left for clarity, but represent the same bins
as the 20~per cent threshold results.
Number counts for the AzTEC/COSMOS and AzTEC/GOODS-N 
surveys have been re-calculated using the final
AzTEC/SHADES prior (solid curve), while applying the methods 
of this paper to the data sets of \citet{scott08} and
\citet{perera08}, respectively, and
are calculated for slightly different bins (i.e. not shifted) 
for improved clarity.
Error bars represent 68~per cent confidence intervals.  
Bin centres are weighted by the assumed prior (solid curve).
The thick horizontal dashed line represents the `survey limit', defined
as the source density that will Poisson deviate to zero sources 
32.7~per cent of the time in a survey this size.
The dot, dash, dash-dot, and dash-dot-dot-dot curves represent the 
predictions of \citet{rowan-robinson08}, \citet{granato04}, 
\citet{vankampen05}, and \citet{baugh05}, respectively.
The thick and thin dotted curves represent models with 
high-redshift formation limits of 
$z_{\textrm{f}} = 4$ and $z_{\textrm{f}} = 5$, respectively. 
}
\label{fig:dnc}
\end{figure}

The combined AzTEC/SHADES differential number counts using 
5~per cent (open circles) and 20~per cent (filled circles) thresholds 
are compared in Fig.~\ref{fig:dnc}.  
The two results are nearly identical at high fluxes, but differ
slightly in the lower flux bins.  
The variation at low flux is not surprising 
given that the 180 additional source candidates 
being considered when using the softer 20~per cent threshold 
are all relatively low in flux,
thus providing significantly more data in the lower flux bins.
All AzTEC/SHADES number count uncertainties
represent the 68~per cent confidence interval derived from the  
distribution of bootstrap iterations.
All uncertainties assume a spatially random distribution 
of sources and, therefore, do not account for the effects of
cosmic variance/clustering.  
The differential number counts data points are strongly 
correlated, as described in Appendix~\ref{app:a}.

Fig.~\ref{fig:inc} 
presents the integral source counts, $N({>}\,S)$,
for both the LH and SXDF surveys
using the 20~per cent null-threshold.  
Unlike the finite-bin differential counts measurements, 
the integral counts are threshold measurements
(i.e. number of sources greater than flux, $S$) and
can be derived at continuous values of flux.
Therefore, the final combined AzTEC/SHADES results
are depicted as 
a continuous 68~per cent confidence region.    
The combined differential and integral number counts 
are also given 
in Table~\ref{tab:counts}, 
with integral counts listed at integer flux limits.

\begin{figure}
\epsfig{file=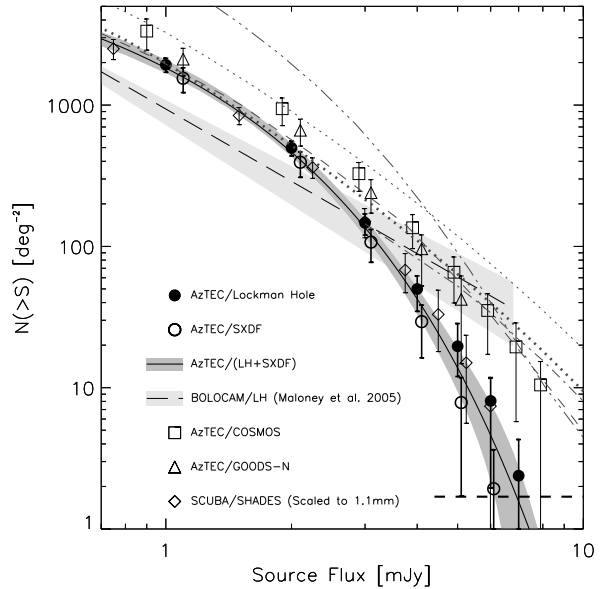, width=1.0\hsize,angle=0}
\caption{Integrated number counts for the AzTEC/LH 
and AzTEC/SXDF 
surveys with 68~per cent confidence intervals.
Their combined constraint on the average blank-field number 
counts is shown as a continuous 68~per cent confidence region.
A 3-parameter Schechter fit to the combined differential 
counts (Fig.~\ref{fig:dnc}) is shown as the solid curve.  
Results of other 1.1--mm surveys are plotted for comparison, 
with the AzTEC/GOODS-N and AzTEC/COSMOS results being recalculated
using the AzTEC/SHADES prior and a 20~per cent null-threshold. 
The SCUBA/SHADES results \citep{coppin06} 
are scaled to 1.1\,mm using the combined
fits of Section~\ref{ssec:nc_fits}.
Discrete integrated number counts data points are 
calculated at 
varying flux values (i.e. not shifted)  
for increased clarity.
Model predictions are plotted as described in Fig.~\ref{fig:dnc}.
}
\label{fig:inc}
\end{figure}

\begin{table}
\caption{\label{tab:counts} 
AzTEC/SHADES differential and integral number counts, calculated
as described in the text.  
The differential number counts flux bins are 1-mJy wide 
with effective bin centres (first column) weighted 
according to the assumed prior.
Correlations amongst data points are described in Appendix~\ref{app:a}.  
}
\begin{center}
\begin{tabular}{cccc}
\hline
Flux Density & dN/dS                   & Flux Density       &  N($>$S)    \\
(mJy)        & ($\textrm{mJy}^{-1}\textrm{deg}^{-2}$)    & (mJy)              &  ($\textrm{deg}^{-2}$)\\
\hline
1.38 \T \B & $1410^{+170}_{-180}$ & 1.0 & $1890^{+190}_{-190}$ \\
2.40 \T \B & $345^{+44}_{-48}$ & 2.0 & $481^{+49}_{-51}$ \\
3.40 \T \B & $94^{+15}_{-18}$ & 3.0 & $136^{+18}_{-20}$ \\
4.41 \T \B & $28^{+7}_{-8}$ & 4.0 & $42^{+9}_{-9}$ \\
5.41 \T \B & $9.2^{+3.4}_{-4.6}$ & 5.0 & $14^{+4}_{-5}$ \\
6.41 \T \B & $3.6^{+1.7}_{-2.8}$ & 6.0 & $4.9^{+2.5}_{-3.0}$ \\
7.41 \T \B & $1.2^{+0.9}_{-1.2}$ & 7.0 & $1.3^{+0.5}_{-1.3}$ \\
\hline
\end{tabular}
\end{center}
\end{table}

\subsection{\label{ssec:nc_simulations}Simulations and tests}

We test for biases and systematics in these techniques by applying the
same number count algorithms to simulated maps of 
model source populations.
Simulated maps are constructed as described in 
Section~\ref{ssec:flux_corrections} and we test a range of 
input model populations motivated by past 1.1--mm and 850--$\micron$
surveys. 
 
\begin{figure}
\epsfig{file=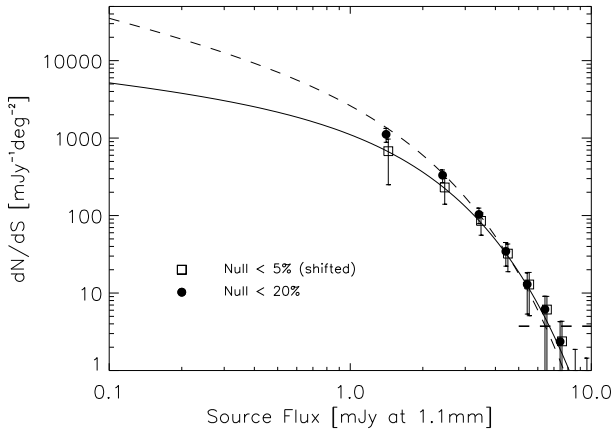, width=1.0\hsize,angle=0}
\caption{
An example of the increase in sensitivity and reduction of bias in 
the low-flux number counts
when using the 20~per cent null threshold in favour of the 5~per cent threshold.  
Here, the differential number counts of the AzTEC Lockman Hole 
survey have been re-calculated (symbols) 
when assuming a significantly 
different (and poor) prior that predicts a much lower number of 
faint sources (solid curve).
For comparison, the dashed curve represents the prior used 
throughout the rest of this paper, which is very near
to our best estimate of the true AzTEC/Lockman-Hole 
number counts. 
The various priors are displayed down to 0.1~mJy to
highlight the differences between the models and do not represent a 
measurement at these fluxes. 
The open symbols have been artificially displaced by $+$5~per cent along
the x-axis for improved clarity.  
}
\label{fig:simnc_test}
\end{figure}

Our simulations show that
the number counts estimates for 
any flux bin
can be significantly biased towards the 
assumed value in the Bayesian prior, 
particularly if that bin is poorly sampled
by the catalogue of source PFDs used to construct the number counts.  
We significantly reduce this bias in the lower flux bins
by extending the sampled catalogue to include fainter source candidates 
with $P(S_\textrm{i} < 0)$ values up to 20~per cent, thus providing more 
data in these otherwise poorly sampled bins. 
This is shown through example in Fig.~\ref{fig:simnc_test}.
Although significant bias to the chosen prior can still be
seen in the lowest flux bin (1-2\,mJy), this bin 
is still very sensitive to the `true' population. 
Therefore, by iteratively adjusting the prior 
based on the results (Section~\ref{ssec:mapmaking}), 
we find that the bulk of this bias can be removed. 
This general result is also supported through full simulations
with a precisely known input population.
As expected, the results based solely on the brightest source
candidates (null-threshold of 5~per cent; open squares) are more severely 
biased by the assumed prior at low fluxes.

The primary concerns when considering low-significance 
sources (e.g. $S/N = 3.0$) are: (a) false detections (noise peaks);
and (b) source confusion (significant contribution from 
multiple sources in each measurement). 
However, false detections are inherently accounted for by having 
accurate PFDs at the intrinsic fluxes being probed, and  
our simulations show that confusion does not play a significant role at 
fluxes $S_\textrm{i} > 1$\,mJy, based on an extrapolation
of measured SMG number counts 
(e.g. \citealp{coppin06}; \citealp{perera08}; this paper)
and the AzTEC/JCMT beamsize (FWHM $= 18$\,arcsec).   
Using the fitted results of Section~\ref{ssec:nc_fits},
the traditional rule of thumb confusion `limit' of 1 source 
per 30 beams \citep[$\Omega_{\textrm{beam}} = \pi\sigma_{\textrm{beam}}^2$; e.g.][]{hogg01}
is $\sim 0.8$\,mJy for AzTEC/JCMT 1.1\,mm data
and is below the most likely intrinsic fluxes
of the individual sources considered here.  
Most importantly, our simulations find
no significant systematics or biases
between the input and output number counts of the constructed maps, 
thus confirming that neither
of the above concerns present a problem for the AzTEC/SHADES 
results as given.

\subsection{\label{ssec:nc_fits}Parametric fits}

For simulation and modelling of the SMG population, 
it is often useful to have
a functional form for the number counts result.  
We fit the AzTEC/SHADES differential number counts to the 
3-parameter Schechter function 
\be
\label{eq:schechter}
{dN \over dS} = {N^\prime} \left({S \over S^\prime}\right)^{\alpha+1} e^{-S/S^\prime},
\ee
using Levenberg-Marquardt minimisation. 
We convert the normalisation parameter $N^\prime$ to the 
more easily interpreted $N_{3\textrm{mJy}}$ 
(the differential counts at $S_{1100} = 3$\,mJy) 
using the relation

\be 
\label{eq:3mjy_conversion}
N_{\text{3mJy}} = N^\prime \left({\text{3\,mJy} \over S^\prime}\right)^{\alpha+1} e^{-3\,\text{mJy}/S^\prime}.
\ee
  
The best-fit AzTEC/SHADES parameters are listed in Table~\ref{tab:fits}. 
The table also includes the results of a combined analysis of 
the currently 
available AzTEC `blank-field' surveys  
AzTEC/SHADES and AzTEC/GOODS-N; however, the addition
of the relatively small GOODS-N survey provides only 
a slight increase in the constraint of the average SMG population.
These results are relatively insensitive 
to the Schechter parameter $\alpha$,
which is strongly anti-correlated to, and somewhat degenerate with, 
the parameter $S^{\prime}$  
in the flux range sampled ($S_\textrm{i} > 1$\,mJy).
Therefore, we find that the AzTEC/SHADES number counts are
nearly as well described by a Schechter function with the $\alpha$ parameter
fixed to a reasonable value that is consistent with previous 
data sets \citep[e.g. $\alpha = -2$;][]{coppin06}. 

In previous incarnations of the bootstrap sampling method outlined 
in Section~\ref{ssec:nc_algorithm} \citep{coppin06,perera08,austermann08}, 
formal fits to the differential 
number counts resulted in unrealistically low $\chi^2$ values,
due to an underestimate of the correlations between bins.
We have now improved the algorithm for calculating the correlation 
matrix, which is described in Appendix~\ref{app:a}.
However, the large correlations amongst the 1\,mJy wide 
AzTEC/SHADES flux bins
lead to a level of degeneracy that significantly complicates 
the application of typical fitting algorithms
that incorporate the covariance matrix.  

\begin{figure}
\epsfig{file=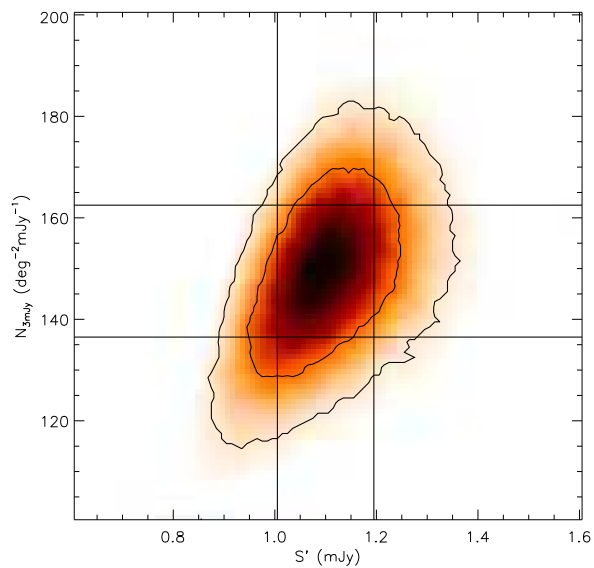, width=1.0\hsize,angle=0}
\caption{Example distribution of best fit results of each iteration 
of bootstrapped AzTEC/SHADES number counts.  
Fits are to \Eq\ref{eq:schechter} with the Schechter parameter
$\alpha$ fixed to $-2$.  Contours represent the 68 and 95~per cent confidence
regions.  Vertical and horizontal lines represent the marginalised
68~per cent confidence intervals of $S^\prime$ and $N_{\textrm{3mJy}}$, respectively.
}
\label{fig:extended_bootstrap}
\end{figure}

We avoid such complications in the derivation of best-fit statistics
by implementing a bootstrap sampling method of  
parameter uncertainty estimation 
that is 
similar to what is used in the error estimation for the individual 
number count data points (Section~\ref{ssec:nc_algorithm}). 
In this method, parameter space is explored by calculating 
best-fit parameters for \textit{each} of the 
100,000 number count bootstrap iterations.
Fig.~\ref{fig:extended_bootstrap} shows the resulting
parameter space for a two-parameter fit to the AzTEC/SHADES
results using 
\Eq\ref{eq:schechter} with Schechter parameter $\alpha$ fixed to a 
value of $-2$.  
Marginalised 68~per cent confidence intervals are used 
for the parameter uncertainties presented in 
Table~\ref{tab:fits}. .
We find that this alternative approach gives results that are comparable to 
that of formal fits, while providing a better characterisation of
the true parameter probability distributions
by avoiding assumptions of Gaussian distributed uncertainty
in the fitted parameter and number count errors.
Since an explicit flux value is chosen for each source
upon an individual iteration of the bootstrap
(i.e. a single flux is chosen from the source's PFD), 
the
number counts found by each realisation have flux bins that are 
effectively independent; therefore, this method provides 
a direct exploration of parameter space 
without necessitating an explicit 
calculation of the bin-to-bin correlations that exist amongst the final averaged
results of Table~\ref{tab:counts}.

\begin{table*}
\caption{\label{tab:fits} 
Parametric fits to differential number counts 
of Table~\ref{tab:counts} using \Eq\ref{eq:schechter}.
Uncertainties of AzTEC-only fits represent the marginalised 
68~per cent confidence intervals derived from the distribution of bootstrap 
iterations (Section~\ref{ssec:nc_fits}).
The Schechter parameter $\alpha$ is given in Column 4, and is held 
constant for fit results given without a confidence interval.  
The quantity $\alpha_{\textrm{dust}}$ is a free parameter 
representing the spectral index inferred 
by the simultaneous fit of the
AzTEC/SHADES (1100--$\micron$) and SCUBA/SHADES (850--$\micron$) 
results, as described in the text.
Uncertainties of the combined AzTEC and SCUBA fit are the formal 
1$\sigma$ parameter errors when using
Levenberg-Marquardt minimisation.
The additional negative correction listed for $\alpha_{\textrm{dust}}$ 
represents an estimated correction for the systematic error induced by 
the SCUBA/SHADES choice of prior, as described in the text.  
All parameter values are for number counts at 1100--$\micron$.
}
\begin{center}
\begin{tabular}{lcccc}
\hline
Data Set & $S^\prime$    & $N_{\textrm{3mJy}}$      & $\alpha$ & $\alpha_{\textrm{dust}}$ \\
         & ($\textrm{mJy}$)       & ($\textrm{deg}^{-2}\textrm{mJy}^{-1}$)  &	    &          \\
\hline
AzTEC/SHADES                &  $1.11^{+0.09}_{-0.09}$ \T \B & $153^{+9}_{-17}$ & $-2$ & -- \\
AzTEC/SHADES                & $1.03^{+0.11}_{-0.43}$  \T \B & $158^{+16}_{-21}$  & $-1.8^{+1.1}_{-0.6}$  & -- \\
AzTEC/SHADES $+$ AzTEC/GOODS-N & $0.96^{+0.25}_{-0.33}$  \T \B & $170^{+15}_{-20}$  & $-1.56^{+0.65}_{-0.9}$  & -- \\
\hline
Two-Frequency Fits &  & & & \\
\hline
AzTEC/SHADES + SCUBA/SHADES & $1.15\pm0.07$ & $153\pm12$ \T \B & $-2$             & $(3.81-0.24)\pm0.17$       \\
AzTEC/SHADES + SCUBA/SHADES & $1.04\pm0.21$ & $157\pm15$ \T \B & $-1.75\pm0.48$ & $(3.83-0.25)\pm0.17$ \\
\hline
\end{tabular}
\end{center}
\end{table*}

Table~\ref{tab:fits} also includes the results of
simultaneous fits to the 
AzTEC/SHADES and published SCUBA/SHADES \citep{coppin06} results, 
where we have assumed the two surveys are sampling the 
same source population and that 
the number counts of the two bands are consistent 
within an average scaling of flux density. 
These fits are accomplished through the introduction 
of a free spectral index parameter, $\alpha_{\textrm{dust}}$, 
which we have defined to reflect the average flux ratio between the two observing 
bands through the relation
\be
\label{eq:flux_ratio}
{S_{850} \over S_{1100}} = \left({\lambda_{850} \over \lambda_{1100}}\right)^{-\alpha_{\text{dust}}},
\ee
where $\lambda_{850}$ and $\lambda_{1100}$ represent the effective centre wavelengths of
AzTEC and SCUBA, respectively \citep[see also][]{perera08}.  
The combined SHADES fit, assuming the nominal band centres
of 850\,$\micron$ and 1100\,$\micron$ and using Levenberg-Marquardt minimisation, 
gives $\alpha_{\textrm{dust}} = 3.81\pm0.17$ 
($S_{850}/S_{1100} = 2.67\pm0.12$).  
The quoted uncertainties do not include systematic errors
due to spectral differences between the SMGs and flux calibrators,
which is expected to be smaller than the formal 1$\sigma$ uncertainty given,
or any systematic calibration errors between the data sets
(the 1$\sigma$ uncertainty of the formal fit is equivalent in size to a $\sim$5~per cent
systematic error in the measured flux ratios).
For optically thin thermal dust emission in the Rayleigh-Jeans limit
($\lambda \gg hc/kT$),
$\alpha_{\text{dust}}$ represents the 
dust emissivity index ($\alpha_{\text{dust}} = 2 + \beta$);
however, the Rayleigh-Jeans approximation is not strictly applicable
at these wavelengths due to the expected temperature 
\citep[$T_{\textrm{d}} \sim 35$\,K; e.g.][]{chapman05,kovacs06,coppin08}
and redshift \citep[$\langle z \rangle \sim 2.2$; e.g.][]{chapman05} of the typical SMG.

We have tested for various other  
systematics between the two instruments' data sets
by recalculating the AzTEC number counts under conditions 
and assumptions closely matching those 
existing for the SCUBA/SHADES analysis \citep{coppin06}.
Since the AzTEC/SHADES survey includes additional mapped area
not covered by SCUBA, the comparison of the two surveys 
may be susceptible to cosmic variance on large scales 
(i.e. $\gsim 0.1\,$deg$^2$).
However, we find that there are no significant differences in the results when
restricting the AzTEC analysis to only those regions 
covered by SCUBA.
We also find no significant differences when applying the same 
5~per cent null-threshold (Section~\ref{ssec:nc_algorithm})
used in the SCUBA/SHADES analysis.  
The SCUBA analysis lacks a correction for the bias to 
peak map locations (Section~\ref{ssec:flux_corrections}), 
however, their use of the conservative 5~per cent null-threshold 
should keep this bias relatively small 
and it is expected to have no significant effect on the 
resultant number counts.

Finally, we note that the SCUBA analysis uses an external 
Bayesian prior that was based on 850--$\micron$ results of the 
Hubble Deep Field North \citep{borys03}, as opposed to the
self-consistent iterative prior used in this paper.  
This prior represents a slight overabundance of bright SMGs when
compared to the SCUBA/SHADES number counts, probably resulting 
in a small, but systematic, overestimate of the SCUBA/SHADES counts.  
Although we cannot use the exact same prior as SCUBA without 
inherently assuming an 850\,$\micron$/1100\,$\micron$ scaling relation
(e.g. a value of $\alpha_{\textrm{dust}}$), we can adopt a similar prior
that assumes the results of an 1100--$\micron$ survey of the
same approximate field \citep[AzTEC/GOODS-N;][]{perera08}.
We recalculate the AzTEC/SHADES counts with this prior,
as well as other matched systematics (5~per cent null-threshold, no 
correction for peak bias) to 
re-estimate $\alpha_{\textrm{dust}}$.
These values are compared to the 
previous fits to determine the systematic error estimates
given in Table~\ref{tab:fits} and result in a final corrected
scaling index of $\alpha_{\textrm{dust}} \approx 3.6\pm0.2$.
As discussed in Section~\ref{sec:discussion}, this value 
of $\alpha_{\textrm{dust}}$ is significantly larger than 
that inferred by 
current measurements of the SMG redshift distribution \citep{chapman05}
and the SEDs of local starbursts \citep{dunne01}, 
as well as direct measurements of SMG flux ratios \citep{kovacs06,coppin08,greve08,chapin09}. 
The lower $S_{850}/S_{1100}$ flux ratios detected through a direct comparison of 
SMGs detected in the GOODS-N field by both AzTEC and SCUBA \citep{chapin09} indicate that 
our relatively high inferred flux ratio may be limited to the comparison of number counts and
not due to systematic calibration errors between the two instruments.
This suggests that the differences 
between the AzTEC and SCUBA number counts analyses
and/or our assumptions of the source population
(i.e. uniform flux ratio and the two wavebands track the same SMG population)
lead to significant systematic errors in the inferred flux ratio 
at this level of sensitivity. 
Analysis of the 850$\micron$/1100$\micron$ 
flux density ratios of individual SHADES sources is deferred to 
Negrello et al. (in prep.).


\section{Discussion}
\label{sec:discussion}

\subsection{\label{ssec:compare_1.1mm}Comparison of 1.1\,mm surveys}

To appreciate the contribution of AzTEC/SHADES
to our understanding of the SMG population, it must be compared
to previous SMG surveys.
The AzTEC/SHADES integral number counts 
are in strong disagreement with 
the parametric results derived from a
fluctuation analysis of the 1.1--mm BOLOCAM Lockman Hole 
survey \citep[dashed line in Fig.~\ref{fig:inc},][]{maloney05}.
The BOLOCAM survey is significantly smaller and shallower
than this AzTEC/SHADES survey 
and consequently contains fewer sources and is more
susceptible to sample variance.    
Furthermore, the BOLOCAM fluctuation analysis is
likely to be skewed by their requirement that the source population be
well described by a single power law, 
which diverges at zero flux and 
has since been shown to poorly describe
the SMG population over a wide range of flux densities
(e.g. \citealp{scott06}; \citealp{coppin06}; this data set).  
Therefore, the
BOLOCAM/LH single power-law result may represent a compromise 
between the relatively steep drop in SMG number
counts at high flux density and the inevitably more moderate slope 
at the faint end.  

Fig.~\ref{fig:inc} also shows the integral number counts
for the individual AzTEC/LH (filled circles) and AzTEC/SXDF
(open circles) fields at specific flux density limits.  
The two fields' number counts are consistent within their 
respective uncertainties; however, 
the overall trend suggests that the AzTEC/LH 
field is rich in bright ($S_{1100} \gsim 4$\,mJy) sources
relative to AzTEC/SXDF.  
This difference of bright AzTEC source counts 
is consistent with the differences seen between the 
regions of LH and SXDF surveyed by SCUBA at 850\,$\micron$ 
\citep{coppin06}.

The effects of cosmic variance appear more prevalent when comparing the
results of this survey to the $0.15$\,deg$^2$
AzTEC/COSMOS survey \citep{scott08}, 
which targeted a region with significant structure,
as traced by the optical/IR galaxy population at $z \lsim 1.1$
\citep{scoville07b}.  
The average blank-field number counts of AzTEC/SHADES
confirm the significant over-density of bright 1.1--mm sources
in the AzTEC/COSMOS region first reported in \citet{austermann08},
who conclude that the observed 
over-density is probably due to gravitational 
lensing by foreground ($z \lsim 1.1$) structure
(comparisons between SHADES sources and other
populations/structure will be explored in future SHADES papers).
We have recalculated the AzTEC/COSMOS number counts 
using the AzTEC/SHADES prior and a 20~per cent 
null-threshold (Fig.~\ref{fig:inc}), 
affirming that the AzTEC/COSMOS over-density is significant 
regardless of the chosen prior.  

We similarly find that the AzTEC/GOODS-N region is relatively 
rich in 1.1--mm sources compared to the much larger
AzTEC/SHADES survey.  This is consistent with the relative abundances 
found in the comparable 850--$\micron$ surveys of 
GOODS-N \citep{borys03} and SHADES \citep{coppin06}.
The higher number counts of AzTEC/GOODS-N may be due 
to sample and/or cosmic variance on the scale of the 
GOODS-N map (0.068\,deg$^2$ to $\sim1$\,mJy), which 
can be exemplified by moving a box the size of AzTEC/GOODS-N 
to different locations within the well-covered, and similar depth,
regions of the AzTEC/LH map (e.g. dash-dotted rectangle 
in Fig.~\ref{fig:map_a}).
This simple exercise shows that the total number of 
source candidates within the
GOODS-N sized box can change by a factor of $\sim$\,2
for any of the source definitions explored here 
(i.e. null-threshold 5--20~per cent). 
The relatively high number of bright SMGs found in 
GOODS-N may be due, in part, to potential 
high-redshift structures in the GOODS-N field \citep{chapman08,daddi08}.  

To better quantify the empirical variations across fields, we turn to the 
formalism of \citet{efstathiou90}. 
Taking the three `blank-fields' considered here 
(AzTEC/LH, AzTEC/SXDF, AzTEC/GOODS-N) as independent 
`cells' and using Equations 9 \& 5 from \citet{efstathiou90}, we can calculate 
$\sigma^2$ and $\textrm{Var}(\sigma^2)$, where $\sigma^2$ 
is defined by Equation 1b of the same paper 
and represents the variation from cell to cell, 
or in this case field to field. 
The variations in field size and completeness are taken into 
account as per the `counts in cells' formalism of Efstathiou (1990).
For $\textrm{Var}(\sigma^2)$ 
we assume zero variance ($\sigma^2=0$) such that the 
significance of the measurement is 
essentially the `detection' significance of some field to field variation.

Number count data at the lowest fluxes, $S_{1100}<2$ mJy, are omitted 
from this analysis due to their sensitivity to the assumed
Bayesian prior,
which is held constant for all fields (Section~\ref{sec:counts}). 
Any bias to the prior would result 
in measured field-to-field variations that are systematically 
lower than that expected from Poisson statistics.  
This bias is relatively small for the remaining data ($S_{1100} > 2$\,mJy)
and, to the extent that it is present, 
would act to make the fields' measured number counts systematically less 
varied and our empirical variance measurements 
conservatively low. We combine the counts such that we 
test the variations in 
two relatively well-sampled flux bins:  
2--4\,mJy, and $>4$\,mJy.

Table~\ref{tab:fluctuations} contains the results of this
variance analysis. 
For the AzTEC `blank-fields' alone, no significant detection of 
variance is found.
The 2--4\,mJy flux bin has a $\sigma^2$ that is 
strongly negative, 
indicating that the measured variance of that bin is
significantly less than that expected from Poisson statistics, 
while the $>4$\,mJy flux bin has a variance consistent with $\sigma^2=0$.
The former may be a consequence 
of the fact that AzTEC/LH and AzTEC/SXDF have coincidentally similar number
counts compared to their formal uncertainty in the 2--4\,mJy range.

The results when including the AzTEC/COSMOS data as an additional cell in 
the analysis are also given in Table \ref{tab:fluctuations}. 
It can be seen that for the brightest sources $S_{1100}>4$ mJy some 
variance is detected at the 3.8$\sigma$ level.
This further confirms the significant over-density 
of bright 1.1--mm sources in the AzTEC/COSMOS region.

\begin{table}
\caption{Statistical significance of the variations in the number 
counts between the three fields considered here (LH, SXDF, GOODS-N), 
as well as the COSMOS field}
\label{tab:fluctuations}
\begin{center}
\begin{tabular}{lllll}
Flux  & $\sigma^2$ & $\Delta(\sigma^2)$& Significance                        &  N$^\prime$ \\
(mJy) &            &                   & $\left(\frac{\sigma^2}{\Delta\sigma^2}\right)$ & ($deg^{-2}$) \\

\hline
\multicolumn{5}{l}{LH, SXDF, GOODS-N only}\\
\hline
2--4 & $-8.9\times10^{-3}$ & 2.3$\times10^{-2}$ & -- & 464.5\\
$>4$ & 7.9$\times10^{-2}$ & 8.2$\times10^{-2}$ & 0.96 & 50.3\\
\hline
\multicolumn{5}{l}{LH, SXDF, GOODS-N and COSMOS}\\
\hline
2--4 & 1.1$\times10^{-2}$ & 2.3$\times10^{-2}$ & 0.48 & 489.9\\
$>4$ & 2.4$\times10^{-1}$ & 6.4$\times10^{-2}$ & 3.8 & 62.2\\
\hline
\end{tabular}
\end{center}
Notes: Negative $\sigma^2$ indicates the measured variance is less than expected from Poisson statistics.
\end{table} 

Interestingly, with the exception of the $>4$ mJy sources across all 4 fields, 
these results are fairly consistent with what would 
be expected from consideration 
of the expected form of the correlation function. 
The expected variance can be calculated by integration 
of the correlation function
$$  \sigma^2 =\int\int_V \xi(r)\,dV \,dV,$$ 
where the integral is calculated over a 
volume, $V$, defined as  
a truncated cone of solid angle $\Omega$ over the redshift range $2<z<3$. 
We assume a correlation function of the form $\xi(r)=(r/r_0)^{-\gamma}$, with 
$r_0=5h^{-1}{\rm Mpc}$ and $\gamma=1.8$, which are 
typical for local galaxy populations. 
Assuming field sizes in the range 0.07--0.37\,deg$^2$. 
gives a predicted variance of 
between 0.017 and 0.008, respectively. 
The measured variance for the number density of 
$S_{1100} > 4$\,mJy sources across all 4 fields is in 
significant excess of that 
predicted under the above assumptions.

However it is worth noting that the quoted errors on the 
measured $\sigma^2$ assume no clustering, 
and are therefore underestimates of the true measurement error. 
In addition it is known that the COSMOS field contains a 
significant over-density of 
1.1--mm sources \citep{austermann08}. 
Taken together it is clear that the COSMOS field is simply 
a highly unusual example, 
and the volumes probed by these surveys are not great enough 
to detect a clustering 
signal of bright SMGs through the comparison of 
number counts alone.

\subsection{\label{ssec:alpha_dust}Comparison to 850--$\micron$ counts}

As shown in Fig.~\ref{fig:inc}, the SCUBA 850--$\micron$ 
and AzTEC 1100--$\micron$ SHADES counts are consistent within a uniform 
scaling of flux density (Section~\ref{ssec:nc_fits}).  
Under the assumption that AzTEC and SCUBA are sampling the same source population
(ignoring selection effects),
$\alpha_{\textrm{dust}}$ represents 
a power law approximation to the 
average redshifted SMG SED at observed wavelengths of $\sim1$\,mm.
The relatively steep 850\,$\micron$/1100\,$\micron$ spectral index 
derived from the SMG populations of SHADES 
(after an approximate correction of systematics due to chosen priors; 
see Section~\ref{ssec:nc_fits}), 
$\alpha_{\textrm{dust}} \approx 3.6\pm0.2$, 
is roughly consistent with the 450\,$\micron$/850\,$\micron$
spectral index, $\alpha_{\textrm{dust}} \approx$ 3.6--3.7, 
found by the SCUBA Local Universe Galaxy Survey (SLUGS)
of IR bright galaxies in the local
Universe \citep{dunne01} after correcting for 
an average CO(3--2) contamination of 25~per cent in the 850--$\micron$ band
at $z=0$ \citep{seaquist04}.
However, the 450\,$\micron$ measurements from SLUGS are already shortward
of the Rayleigh-Jeans limit (but longward of the peak) in the local Universe; 
particularly as local galaxy SEDs require two or more dust temperature 
components, with the cooler component being $\sim20$\,K.
For a population of SMGs residing at the typical redshift of 
$z \sim2$, the observed 850--$\micron$ SCUBA band is sampling 
a rest-frame wavelength of $\sim$\,280\,$\micron$.  
To produce a similar $\alpha_{\textrm{dust}}$ to the SLUGS
galaxies in the local Universe, a much hotter temperature is required 
for the SED ($T \ge 50$\,K with $\beta = 2$).
Alternatively, these SMGs could have similar SEDs to the local galaxies but 
reside at lower redshifts ($z \lsim 1$), although this would be inconsistent
with measured SMG redshift distributions \citep{chapman05}.  
The inferred sub-mm/mm flux ratio is 
high compared to the model predictions of \citet{swinbank08} and
at odds with measurements of the flux ratios
$S_{350}/S_{850}$ \citep{kovacs06,coppin08}, 
$S_{850}/S_{1100}$ \citep{chapin09}, 
and $S_{850}/S_{1200}$ \citep{greve08}, 
which are all more consistent 
with the SLUGS SEDs for $z\sim2$.
In addition, the existence of a population of 
sub-millimetre drop-outs \citep[SDOs; e.g.][]{greve08} 
-- sources with a combination of high-redshift and/or low 
dust temperature such that the 850--$\micron$ band 
samples near, or shortward, of the peak emission --
would act to \textit{lower} the average value $\alpha_{\textrm{dust}}$
for millimetre detected sources.  

It thus appears that our estimate of $\alpha_{\textrm{dust}}$ 
is systematically large given the
expectation that $\beta \lsim 2$ \citep[][and references therein]{dunne01}
and that the majority of our sources 
are unlikely to be fully in the Rayleigh-Jeans limit.
This bias may be indicative of 
further systematics in the SCUBA/SHADES choice of prior
(Section~\ref{ssec:nc_fits}), potentially insufficient de-boosting of 
low $S/N$ SCUBA detections (as suggested in a direct comparison
of sources detected by both AzTEC and SCUBA in the GOODS-N field; \citealp{chapin09}), 
or that selection bias somehow results 
in a systematic \textit{increase} in the value of 
$\alpha_{\textrm{dust}}$ inferred from SMG number counts
when assuming 850\,$\micron$ and 1100\,$\micron$ 
sample the same approximate source population. 
A straight comparison of the AzTEC and SCUBA SHADES maps (Negrello et al., in prep.)
will provide a more direct measure of 
$\alpha_{\textrm{dust}}$ that is based on individual sources and fluxes in the maps
and search for evidence of SDOs in the SHADES fields.

\subsection{\label{ssec:discussion_models}Predictions from models}

Finally, we compare the AzTEC/SHADES number counts to those predicted 
at 1100\,$\micron$ by
various IR/sub-mm formation and evolution models
in figs.~\ref{fig:dnc} and \ref{fig:inc}.
The predictions of the IR/sub-mm evolution models of \citet{rowan-robinson08}
are shown for
high-redshift formation limits of $z_{\textrm{f}} = 4$ and $z_{\textrm{f}} = 5$.
The AzTEC/SHADES number counts agree with the $z_{\textrm{f}}=4$ model at 
fluxes $S_{1100} \lsim 4$\,mJy, but are systematically lower than the predictions
at higher fluxes.   
A semi-analytical model for the joint formation and evolution of spheroids and QSOs 
\citep{granato04,silva05} predicts very similar number counts 
at 1100\,$\micron$. 
These models are in better overall agreement with the high flux number counts seen in the AzTEC/COSMOS
and AzTEC/GOODS-N fields; however, those fields have significant biases and/or limitations,
as discussed above. 
Also compared are the counts predicted 
by the semi-analytical galaxy formation model of 
\citet[][see also \citealp{lacey08,swinbank08}]{baugh05}, 
which systematically over-predicts the number of sources seen in AzTEC/SHADES 
by a factor of 3--4 at all measured fluxes.  
Finally, we compare our results to the early predictions for SHADES  
\citep[][]{vankampen05} -- 
models constrained to the SCUBA 8-mJy (i.e. $S_{1100} \ge 3$\,mJy)
survey \citep{scott02} 
-- which  
forecast a shallower slope in the number counts  
than seen in the AzTEC/SHADES fields.  
Assuming the bright sources are uniformly distributed across the sky,
the AzTEC/SHADES survey suggests that all of 
these models significantly over-predict the number
of intrinsically bright SMGs. 
If, instead, these relatively rare sources are strongly clustered, 
the true all-sky average number density of the brightest SMGs
could be higher (or lower) than indicated by this survey, potentially bridging 
the gap between model and observation.

\section{Conclusions}
\label{sec:conclusions}

AzTEC/SHADES is the largest
extragalactic millimetre-wave survey 
to date, with over  
0.7\,deg$^2$
mapped to depths
of $0.9 < \sigma_{1.1} < 1.7$\,mJy.  
This survey, split between the 
Subaru/XMM-Newton Deep Field and the Lockman Hole,
provides over 100 significant individual detections at 1.1\,mm, 
with most representing 
newly discovered mm-wave sources. 
These maps also provide information on the fainter SMG 
population through the signature of numerous dimmer sources 
that are partially buried in the noise.

Combined with our improved methods for 
number count estimates, AzTEC/SHADES provides the 
tightest available constraints on the average SMG population
in the flux range $1$\,mJy $\lsim S_{1100} \lsim 10$\,mJy.  
In particular, the AzTEC/SHADES results represent a significant 
advance in our knowledge of the blank-field 
population at 1.1\,mm, showing that there are significantly 
lower densities of bright SMGs than that suggested by
smaller 1.1--mm surveys published previously.    
An accurate understanding of the average SMG population is critical for
comparisons to source counts found in biased and/or over-dense regions. 
The AzTEC/SHADES blank-field counts confirm the over-density 
of $S_{1100} > 2$\,mJy sources found in the AzTEC/COSMOS field \citep{austermann08} and
show that the GOODS-N field is also relatively rich in 
bright SMGs, 
thus suggesting that  
cosmic variance can significantly affect the observed number
density of SMGs in mass-biased regions (AzTEC/COSMOS) and/or 
on relatively small scales (AzTEC/GOODS-N; $0.068$\,deg$^2$).
We find that the variance in number counts seen across the 
four available AzTEC/JCMT survey fields (LH, SXDF, COSMOS, GOODS-N)
is significantly larger than that expected from Poisson statistics alone,
particularly at $S_{1100} > 4$\,mJy, 
thus suggesting that bright SMGs may be strongly clustered.

The AzTEC/SHADES results are consistent with the predictions of
the formation and evolution models of \citet{granato04} 
and the $z_{\textrm{f}} \sim 4$ evolution models of \citet{rowan-robinson08}
for blank-field 1.1--mm source counts at $S_{1100} \lsim 4$\,mJy;  
however, these models systematically over-predict the number of AzTEC/SHADES 
sources seen at higher fluxes, although the relative scarcity 
and potential clustering of bright sources 
leaves even this unprecedentedly large SMG survey susceptible 
to the effects of cosmic variance. 
A truly unambiguous characterisation of the $S_{1100} \gsim 6$\,mJy
SMG population will require significantly larger-area surveys at (sub-)mm wavelengths, 
such as those expected to be conducted in the coming year(s) by SCUBA-2  
on the JCMT and AzTEC when mounted on the Large Millimeter Telescope (LMT). 

We find that the SCUBA/SHADES and AzTEC/SHADES number counts 
are consistent within a uniform scaling of flux density. 
Assuming that the 850\,$\micron$ and 1100\,$\micron$ wavebands 
sample the same underlying source population,
this scaling corresponds to an
average source flux ratio of 
$S_{850}/S_{1100} \approx 2.5\pm0.1$, 
once corrected for known systematics between the data sets.
This ratio is significantly larger than that expected for the high-redshift
SMG population and we find that the systematics induced
by small differences in the number count analyses of the
two surveys and the assumption of a uniformly scalable flux density
limit the robustness of the inferred flux ratio.
The S850/S1100 flux ratio is explored further in a direct
comparison of individual sources lying in the overlapping
regions of the SCUBA and AzTEC surveys (Negrello et al.,
in prep.).



\section*{Acknowledgements} 
The authors would like to thank 
J. Karakla,
K. Souccar, 
C. Battersby, 
C. Roberts, 
S. Doyle,
I. Coulson, 
R. Tilanus, 
R. Kackley,  
and the observatory staff at the JCMT who helped make this
survey possible.  
Support for this work was provided in part by 
NSF grant AST 05-40852
and a grant from the Korea Science \& Engineering Foundation (KOSEF)
under a cooperative Astrophysical Research Center of the Structure and 
Evolution of the Cosmos (ARCSEC).
KS was supported in part through the NASA GSFC Cooperative Agreement NNG04G155A.
OA, IRS, RJM and JSD acknowledge support from the Royal Society.
IA and DHH acknowledge partial support by CONACyT from research 
grants 39953-F and 39548-F.
EC, KC, JSD, MH, DS, and AP acknowledge support from NSERC. 
AM, MC, DF, MN, IRS, and AMS acknowledge support from STFC. 
AMS acknowledges support from an RAS Lockyer Fellowship.


\setlength{\bibhang}{2.0em}


\appendix
\section{Correlation Matrix}
\label{app:a}
The bootstrap sampling method of Section~\ref{ssec:nc_algorithm}
induces significant correlation between the final averaged differential 
number count bins (e.g. Table~\ref{tab:counts}) 
through discrete sampling (and consequent binning) of continuous PFDs 
that have significant probability on scales comparable
to, or larger than, the bin size (1\,mJy in this paper).
Previous incarnations of this sampling method 
\citep[e.g.][]{coppin06}
estimated covariance and correlation matrices 
directly from the variation in number count
results seen across the iterations of the bootstrap.  
This sampling method collapses each source's probability distribution 
(PFD; e.g. Fig.~\ref{fig:deboosted}) to a single flux upon each iteration, 
which acts to hide significant correlation amongst the final binned results
by throwing away much of the cross-bin
information contained within the PFD. 
This resulted in severely underestimated bin-to-bin
correlations amongst the differential number counts data, as
was evidenced by the unrealistically low $\chi^2$ 
values of formal fitting \citep{coppin06,perera08,austermann08}.  

\begin{table}
\caption{\label{tab:corr}
Correlation matrix of the AzTEC/SHADES differential count bins
as given in Table~\ref{tab:counts}.
The last row gives the standard deviation of each bin
as determined through the bootstrap sampling method of 
Section~\ref{ssec:nc_algorithm}.
Together, these values can be used to create a covariance matrix;
however, we note that the probability distribution is not strictly 
Gaussian, particularly for the less populated bins, 
as evidenced by the asymmetric uncertainty intervals of
Table~\ref{tab:counts}.
}
\begin{center}
\begin{tabular}{l|ccccccc}
\hline
FLUX & 1.38 & 2.40 & 3.40 & 4.41 & 5.41 & 6.41 & 7.41 \\
(mJy) &     &      &      &      &      &      &      \\
\hline
1.38 & 1.00 & 0.92 & 0.61 & 0.26 & 0.08 & 0.02 & 0.01\\
2.40 & 0.92 & 1.00 & 0.84 & 0.47 & 0.18 & 0.04 & 0.01\\
3.40 & 0.61 & 0.84 & 1.00 & 0.82 & 0.44 & 0.12 & 0.05\\
4.41 & 0.26 & 0.47 & 0.82 & 1.00 & 0.78 & 0.32 & 0.17\\
5.41 & 0.08 & 0.18 & 0.44 & 0.78 & 1.00 & 0.77 & 0.61\\
6.41 & 0.02 & 0.04 & 0.12 & 0.32 & 0.77 & 1.00 & 0.97\\
7.41 & 0.01 & 0.01 & 0.05 & 0.17 & 0.61 & 0.97 & 1.00\\
\hline
\hline
$\sigma$ & 179.77 &  46.40 &  16.69 &   7.55 &   3.98 &   2.38 & 1.33\\
\end{tabular}
\end{center}
\end{table}

We now present an alternative method of calculating the correlation 
matrix which better captures these correlations amongst the 
final differential 
number count bins. 
We begin by integrating the PFD of each source over 
the span of each flux bin.  
These binned probabilities can be summed over
all sources to provide a number counts estimate 
that matches the final averaged results of the 
full bootstrapping method, but without the 
robust uncertainty estimates that the bootstrap
method is designed to provide. 
We apply this alternative number counts estimate to \textit{each}
of the 100,000 unique catalogues produced by the bootstrap.
The Poisson deviation and replacement sampling 
used to produce each catalogue (Section~\ref{ssec:nc_algorithm})
act to perturb this new estimate of the number counts
around the most likely values.
This collection of perturbed number counts 
is then used
to estimate the correlation between the differential number count bins.

We present the resulting correlation matrix for the AzTEC/SHADES
differential results (Section~\ref{sec:counts}) in 
Table~\ref{tab:corr}.  These correlations apply directly to 
the differential results provided in Table~\ref{tab:counts} and
Fig.~\ref{fig:dnc} (20~per cent threshold counts).
The last row of Table~\ref{tab:corr} provides the 
standard deviation of the differential counts of each 
flux bin, as estimated in the bootstrap sampling method. 
These values can be applied to the correlation matrix to 
produce a covariance matrix for the data. 
However, the standard deviation is not an 
ideal representation of the true uncertainty distribution
(Table~\ref{tab:counts}) due to the finite sampling 
of each bin, which results in an asymmetric multinomial 
probability distribution (i.e. non-Gaussian).
As discussed in Section~\ref{ssec:nc_fits}, care must be taken when 
attempting to use this covariance matrix in typical fitting algorithms
due to the high level of degeneracy amongst the bins.  
Larger flux bins could be used to reduce the bin-to-bin correlations;
however, significantly larger bins
would make flux resolution the limiting factor
(with respect to the precision of the AzTEC/SHADES 
differential number counts estimate) 
for most practical applications of the data.

\end{document}